\documentclass[preprint,prc,tightenlines,nofootinbib,superscriptaddress]{revtex4}
\usepackage{amssymb}
\usepackage{amsmath}
\usepackage{graphicx}
\usepackage{bm} 
\usepackage[loose]{subfigure}
\usepackage{xcolor}
\usepackage[squaren]{SIunits}
\usepackage{multirow}
\usepackage{braket}
\usepackage{tensor}
\usepackage{slashed}
\usepackage{array}

\newcommand{\punkt}{\;\text{.}}
\renewcommand{\vec}[1]{\boldsymbol{#1}}
\newcommand{\komma}{\;\text{,}}
\renewcommand{\i}{\mathrm{i}}

\begin{document}

\title{Elastic and inelastic pion-nucleon scattering to fourth order in chiral perturbation theory}

\author{D.~Siemens}
\email[]{dmitrij.siemens@rub.de}
\affiliation{Institut f\"ur Theoretische Physik II, Ruhr-Universit\"at Bochum, D-44780 Bochum, Germany}

\author{V.~Bernard}
\email[]{bernard@ipno.in2p3.fr}
\affiliation{Groupe de Physique Th\'eorique, Institut de Physique
Nucl\'eaire, UMR 8608, CNRS, Univ. Paris-Sud, Universit\'e Paris Saclay,
F-91406 Orsay Cedex, France}

\author{E.~Epelbaum} 
\email[]{evgeny.epelbaum@rub.de}
\affiliation{Institut f\"ur Theoretische Physik II, Ruhr-Universit\"at
  Bochum, D-44780 Bochum, Germany}

\author{A.~M.~Gasparyan}
\email[]{ashot.gasparyan@rub.de}
\affiliation{Institut f\"ur Theoretische Physik II, Ruhr-Universit\"at Bochum, D-44780 Bochum, Germany}
\affiliation{Institute for Theoretical and Experimental Physics, B. Cheremushkinskaya 25, 117218 Moscow, Russia}

\author{H.~Krebs}
\email[]{hermann.krebs@rub.de}
\affiliation{Institut f\"ur Theoretische Physik II, Ruhr-Universit\"at Bochum, D-44780 Bochum, Germany}

\author{Ulf-G.~Mei{\ss}ner}
\email[]{meissner@hiskp.uni-bonn.de}
\affiliation{Helmholtz-Institut f\"ur Strahlen- und
             Kernphysik and Bethe Center for Theoretical Physics, \\
             Universit\"at Bonn,  D--53115 Bonn, Germany}
\affiliation{Institute~for~Advanced~Simulation, Institut~f\"{u}r~Kernphysik and
J\"{u}lich~Center~for~Hadron~Physics, ~Forschungszentrum~J\"{u}lich,
D-52425~J\"{u}lich, Germany}
\affiliation{JARA~-~High~Performance~Computing, Forschungszentrum~J\"{u}lich, 
D-52425 J\"{u}lich,~Germany}

\begin{abstract}
We extend our previous study of elastic pion-nucleon scattering in the
framework of chiral perturbation theory by
performing a combined analysis of the reactions $\pi N \to \pi N$ and
$\pi N \to \pi \pi N$.  The calculation is carried out to fourth order
in the chiral expansion using the heavy baryon approach and the covariant
formulation supplemented with a modified version of the extended
on-mass-shell renormalization scheme. We demonstrate that a combined
fit to experimental data in both channels leads to a reduced amount
of correlations between the low-energy constants. A satisfactory
description of the experimental data in both channels is obtained,
which is further improved upon including tree-level
contributions of the $\Delta$(1232) resonance. We also explore a
possibility of using the empirical information about $\pi N$ subthreshold
parameters obtained recently by means of the Roy-Steiner equations to
stabilize the fits. 
\end{abstract}

\maketitle

\section{Introduction}
In recent years, there has been a revival of interest in theoretical
studies of elastic pion-nucleon scattering. One important milestone is
a new partial-wave analysis in the framework of the Roy-Steiner
equations~\cite{Ditsche:2012fv,Hoferichter:2015dsa}, which incorporates the 
fundamental principles of analyticity, crossing  symmetry and unitarity.
Using the empirical information about  high-energy $\pi N$ and
$\pi\pi$ scattering, the authors of Ref.~\cite{Hoferichter:2015dsa} have performed error
propagation of all input quantities to finally determine pion-nucleon
S- and P-wave phase shifts with quantified uncertainties, see
the review~\cite{Hoferichter:2015hva} for more details.

Considerable progress has also been made toward the understanding of
elastic pion-nucleon scattering in the framework of chiral
perturbation theory ($\chi$PT), an effective field theory of the strong interactions 
that allows one to perform a systematic expansion of low-energy hadronic
observables in powers of the soft scales such as the pion mass $M_\pi$ and/or external
three-momenta of the interacting particles $\vec p_i$. Here and in what follows,
we restrict ourselves to the two-flavor case of the light up- and
down-quarks. Throughout, we work in the isospin limit $m_u=m_d$.
In the single-nucleon sector,  special care is
required to maintain the chiral power counting in the presence of the
nucleon mass $m_N$. This can be achieved using the heavy-baryon scheme or,
alternatively, by exploiting the freedom in the choice of
renormalization conditions in the covariant framework. 

In the heavy-baryon approach, one performs a $1/m_N$ expansion at the
level of the effective Lagrangian
\cite{Jenkins:1990jv,Bernard:1992qa}. As a result, the nucleon mass
only enters the heavy-baryon Lagrangian in the form of
$1/m_N$-corrections to the vertices so that no positive powers of
$m_N$ can emerge when calculating the corresponding Feynman diagrams. 
In the single-nucleon sector, the nucleon mass is counted as a
quantity of the order of the breakdown scale of the chiral expansion
$\Lambda_b$, i.e. $m_N \sim \Lambda_b$. Here and in what follows, we
denote the resulting approach as HB-$\pi$N. In contrast, in
few-nucleon calculations one usually treats the nucleon mass as an
even larger scale via the assignment $m_N \sim \Lambda_b^2/M_\pi$
\cite{Weinberg:1991um,Epelbaum:2008ga}. This approach, which we refer to as HB-NN, leads to a
stronger suppression of relativistic corrections as compared to the
HB-$\pi$N scheme.  Note that in all our estimates we adopt the conservative value of $\Lambda_b \sim
600$~MeV as in Ref.~\cite{Siemens:2016hdi}.

For a covariant formulation of baryon $\chi$PT, the chiral power
counting can be maintained employing the so-called infrared
renormalization scheme \cite{Ellis:1997kc,Becher:1999he} or, alternatively, by using the 
extended on-mass-shell scheme (EOMS)
\cite{Gegelia:1999gf,Fuchs:2003qc}. Here and in what follows, we will
employ the EOMS approach in a slightly different form as compared with
its original formulation. In particular, we require the
$1/m_N$-expansion of our results  to match exactly the heavy-baryon
expansion which can be achieved via performing additional finite
renormalization of the low-energy constants (LECs), see
Ref.~\cite{Siemens:2016hdi} for details. 

In Ref.~\cite{Siemens:2016hdi}, we have studied elastic pion-nucleon scattering to fourth
order in the chiral expansion within both the HB and covariant
formulations. Differently to the previous $\chi$PT studies of this reaction
\cite{Bernard:1992qa,Mojzis:1997tu,Fettes:1998ud, 
Buettiker:1999ap,Fettes:2000gb,Fettes:2000xg,Becher:2001hv,Hoferichter:2009gn,
Gasparyan:2010xz,Alarcon:2012kn,Chen:2012nx}, we have directly used 
the available experimental data taken from the GWU-SAID data base 
\cite{Workman:2012hx} rather than the partial wave analyses such as 
e.g.~the ones performed by the Karlsruhe-Helsinki  \cite{Koch:1985bn}
and GWU (SAID) \cite{Arndt:2006bf} groups
to determine the values of the various
LECs, see also Ref.~\cite{Wendt:2014lja} for a recent work following the same
strategy (where only the HB-NN version of $\chi$PT was considered and no theoretical 
errors were taken into account).
In addition, we have carried out a detailed estimation of
theoretical uncertainties from the truncation of the chiral expansion
by employing  the algorithm formulated in Ref.~\cite{Epelbaum:2014efa}. These two
features have allowed us to directly translate the experimental errors
into the statistical uncertainties of the extracted LECs 
and correlations among them. The predicted phase shifts were found to
be in good agreement with the ones  of Ref.~\cite{Hoferichter:2015dsa}.  
Finally, elastic pion-nucleon scattering
has also been analyzed at the leading one-loop order in a covariant
formulation of $\chi$PT with explicit $\Delta$-resonance degrees of
freedom \cite{Yao:2016vbz}. In that work, the LECs have been determined from fits to
phase shifts determined in the Roy-Steiner equation analysis of
Ref.~\cite{Hoferichter:2015dsa}.

There is a fair number of unknown LECs that need to be
determined from the fit, namely 
$8$ ($13$) LECs at order $Q^3$ ($Q^4$) with $Q \in \{| \vec
p_i |/\Lambda_b, \; M_\pi/\Lambda_b \}$  denoting the expansion
parameter in $\chi$PT. This results in sizeable uncertainties 
and large correlations among some of the LECs.
It is, therefore, desirable to incorporate additional empirical
information when doing the fits in order to further constrain the
values of the LECs. $\chi$PT provides a suitable tool to achieve this
goal as it allows one to apply the same effective Lagrangian to different processes and  
kinematical regions as long as one stays within the applicability
domain of the chiral expansion. 

In the present study we explore  two possibilities for further
constraining the fits. First, we employ the information on the
so-called subthreshold $\pi N$ parameters, which have been  extracted
recently with high accuracy by means  
of the Roy-Steiner equation \cite{Hoferichter:2015hva}. 
Secondly, we perform combined fits of the experimental data in elastic
pion-nucleon scattering and the inelastic reaction $\pi N \to \pi \pi
N$. The corresponding scattering amplitude has been calculated up to the
leading one-loop order (i.e.~$Q^3$) in HB formulation of $\chi$PT in
Refs.~\cite{Bernard:1995gx,Fettes:1999wp}, see Refs.~\cite{Beringer:1992ic,Bernard:1994wf,Olsson:1995iy} for
related earlier studies. Furthermore, single pion production off
nucleons was also analyzed at tree level in the covariant $\chi$PT framework 
with an implicit \cite{Bernard:1997tq} and explicit \cite{Siemens:2014pma} treatment of effects due to
$\Delta$-resonance.  A covariant tree-level investigation including both the $\Delta$ and
the Roper resonances was presented in Ref.~\cite{Jensen:1997em}.
In this work, we extend these calculations by
performing, for the first time, a complete analysis of the reaction $\pi N \to \pi \pi
N$ at the full one-loop order (i.e.~$Q^4$) using both the HB and
covariant formulations of  $\chi$PT. 

Our paper is organized as follows. In section~\ref{sec:SubPar}, we
give the definition of the pion-nucleon subthreshold coefficients
while section~\ref{sec:PionProd} contains the basic definitions and 
formalism for the reaction $\pi N\to\pi\pi N$. The details
of the fitting procedure can be found in section~\ref{sec:fitting-procedure}. The discussion of the 
naturalness of the extracted low-energy constants is presented in
section~\ref{sec:LECs}, where we also discuss the lowest-order
contributions of the  $\Delta$- and Roper-resonances to these LECs. 
Our predictions for various observables are collected in
section~\ref{sec:predictions}, where we also discuss the obtained 
results. Finally, the main results
of our study are summarized in section~\ref{sec:sum}. The appendix
contains explicit expressions for the resonance saturation of  LECs due to the explicit inclusion of lowest-order
$\Delta$(1232)- and Roper-resonance.

\section{Pion-nucleon subthreshold parameters}
\label{sec:SubPar}

As already pointed out in the introduction, this work provides an
extension of the previous analysis of the reaction
$\pi N\to\pi N$ in \cite{Siemens:2016hdi}. In particular, 
we explore the possibility to improve the extraction of the $\pi N$
LECs by incorporating additional constraints from the subthreshold
kinematical region by including the
leading subthreshold parameters in our fitting procedure. 
In the following,  we provide the basic definitions of the  
subthreshold parameters. A detailed discussion of our calculation of
the $\pi N$ scattering amplitude including the definitions of
observables and  kinematics  as well as the details concerning
renormalization up to order $Q^4$ can be found in
Ref.~\cite{Siemens:2016hdi}.   

The $T$-matrix for the process $\pi^a(q) \, N(p) \; \to \; \pi^b(q') \, N^\prime(p^\prime)$
can be conveniently expressed in the form
\begin{equation}
  \label{eq:1}
  \begin{aligned} 
    T^{ba}&=\chi^\dagger_{N^\prime}\left(\delta^{ab} T^+ +\i
      \epsilon^{bac} \tau_c T^-\right)\chi_N \komma\qquad
    T^\pm&=\bar{u}^{(s^\prime)}\left(D^\pm-\frac{1}{4m_N}[\slashed{q}^\prime,\slashed{q}]B^\pm\right)u^{(s)}\komma
  \end{aligned}
\end{equation}
where the amplitudes $D^\pm$ and $B^\pm$ depend on the quantities $t$
and $\nu=(s-u)/4m_N$, with the Mandelstam variables defined as 
\begin{equation}
  \label{eq:17}
     s=(p+q)^2\komma\quad t=(q-q^\prime)^2\komma\quad
  u=(p^\prime-q)^2\komma\quad s+t+u=2m_N^2+2M_\pi^2\punkt
\end{equation}
The subthreshold parameters are defined by an expansion of the amplitudes
in powers of $\nu$ and $t$ via \cite{Hoehler,Becher:2001hv} 
\begin{equation}
  \label{eq:3}
  \begin{aligned}
D^\pm=
\begin{pmatrix}
  1\\\nu
\end{pmatrix}
\sum^{\infty}_{n,m=0} d_{mn} \nu^{2m}t^n+D_{\mathrm{pv}}^\pm
\komma\quad
B^\pm=
\begin{pmatrix}
  \nu\\1
\end{pmatrix}
\sum^{\infty}_{n,m=0} b_{mn} \nu^{2m}t^n+B_{\mathrm{pv}}^\pm\,,
  \end{aligned}
\end{equation}
where $B_{\mathrm{pv}}^\pm$ and $D_{\mathrm{pv}}^\pm$ refer to the
subtracted pseudovector Born-term contributions given by 
\begin{equation}
  \label{eq:4}
  B_{\mathrm{pv}}^\pm=g^2_{\pi NN}\left( \frac{1}{m_N^2-s}\mp \frac{1}{m_N^2-u} \right)-\frac{g_{\pi NN}^2}{2m_N^2}
\begin{pmatrix}
  0\\1
\end{pmatrix}\komma\quad
D_{\mathrm{pv}}^\pm=\frac{g_{\pi NN}^2}{m_N}
\begin{pmatrix}
  0\\1
\end{pmatrix}+\nu   B_{\mathrm{pv}}^\pm\punkt
\end{equation}

\section{The reaction $\pi N\to\pi\pi N$}
\label{sec:PionProd}

We now turn to the reaction $\pi N\to\pi\pi N$ and mainly focus on the
renormalization of the amplitude. To be more precise, we follow the
same procedure as for the elastic channel in Ref.~\cite{Siemens:2016hdi}
and only present in the following the new features appearing in the
pion production process. More details on the studied observables, in
particular the relations to the amplitude, can be found in Ref.~\cite{Siemens:2014pma}.

The $T$-matrix for the reaction
$  \pi^a(q_1) \, N(p) \; \to \; \pi^b(q_2)\, \pi^c(q_3) \, N^\prime(p^\prime)$
can be expressed in terms of four invariant amplitudes
\begin{equation}
  \label{eq:28}
    T^{abc}=\i
    \bar{u}^{(s^\prime)}\gamma_5\left(F_1^{abc}+(\slashed{q}_2+\slashed{q}_3)\tilde
      F_2^{abc}
  +(\slashed{q}_2-\slashed{q}_3)\tilde F_3^{abc}
 +\slashed{q}_1(\slashed{q}_2\slashed{q}_3-\slashed{q}_3\slashed{q}_2)\tilde
 F_4^{abc}\right) u^{(s)} \,,
\end{equation}
which depend on the five Mandelstam variables
\begin{equation}
  \label{eq:25}
    s=(p+q_1)^2\komma\quad s_1=(q_2+p^\prime)^2\komma\quad
  s_2=(q_3+p^\prime)^2\komma\quad t_1=(q_2-q_1)^2\komma\quad
  t_2=(q_3-q_1)^2 \,.
\end{equation}
Notice that in Ref.~\cite{Siemens:2014pma}, a different basis was
chosen to decompose the amplitude. 
The amplitudes $\tilde
      F_i^{abc}$ are related to the ones $F_i^{abc}$ used in Ref.~\cite{Siemens:2014pma} via
\begin{equation}
  \label{eq:29}
  \begin{aligned}
    \tilde F_1 &=F_1\komma\\
    \tilde F_2 &=F_2-\frac{1}{2m_N}(s_1-s_2+t_1-t_2)F_4\komma\\
    \tilde F_3 &=F_3-\frac{1}{2m_N}(4M_\pi^2+m_N^2-s-t_1-t_2)F_4\komma\\
    \tilde F_4 &= -\frac{1}{2m_N}F_4\punkt
  \end{aligned}
\end{equation}
The isospin decomposition of the invariant amplitudes reads
\begin{equation}
  \label{eq:27}
  F_i^{abc}=
\chi_{N^\prime}^\dagger\left(\tau^a\delta^{bc}B_i^1
+\tau^b\delta^{ac}B_i^2+\tau^c\delta^{ab}B_i^3
+\i\epsilon^{abc} B_i^4 \right)\chi_{N} . 
\end{equation}
The basis in Eq.~\eqref{eq:28} is better suited for the
renormalization procedure, because
each spin structure fulfills the power-counting on its own. Like in the
case of $\pi N$-scattering, the individual spin structures
are expanded in small parameters,
\begin{equation}
  \label{eq:30}
  \begin{aligned}
    M_\pi\sim \mathcal{O}(Q^1)\komma\quad s-m_N^2&\sim
    \mathcal{O}(Q^1)\komma\quad s_1-m_N^2\sim
    \mathcal{O}(Q^1)\komma\quad s_2-m_N^2\sim
    \mathcal{O}(Q^1)\komma\\
    t&\sim \mathcal{O}(Q^2)\komma\quad
    t_1\sim \mathcal{O}(Q^2)\komma\quad t_2\sim \mathcal{O}(Q^2)\,,
  \end{aligned}
\end{equation}
which allows one to identify the power-counting breaking terms.
The linear combination $s-s_1-s_2+m_N^2$ counts, 
according to the above rules, as a quantity of order $Q^1$ but
actually starts contributing at order $Q^2$. It is, therefore, advantageous to
express the invariant amplitudes as functions of e.g.~$s_1$,
$s_2$, $t$, $t_1$ and $t_2$. In the following, all LECs should be
understood as renormalized quantities and the explicit shifts used for the
renormalization can be found in Ref.~\cite{Siemens:2016hdi}. 

\begin{figure}[ht]
  \centering
\includegraphics[width=\textwidth]{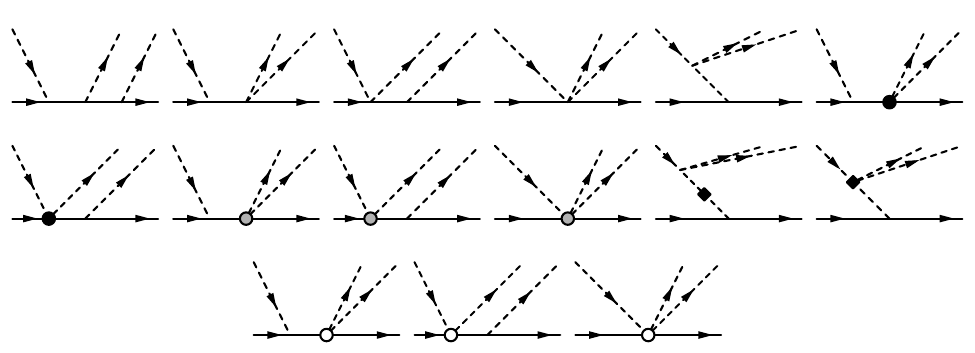}
  \caption{Tree-level graphs contributing to the reaction $\pi N\to\pi\pi N$. 
    The black/gray/white
    blobs denote insertions of the $c_i$/$d_i$/$e_i$- vertices,
    whereas the black diamonds denote insertions of the $l_i$
    vertices. Dashed and solid lines refer to pions and nucleons, respectively. Crossed
    diagrams are not shown.}
  \label{fig:TreeGraphs}
\end{figure}

\begin{figure}[ht]
  \centering
\includegraphics[width=\textwidth]{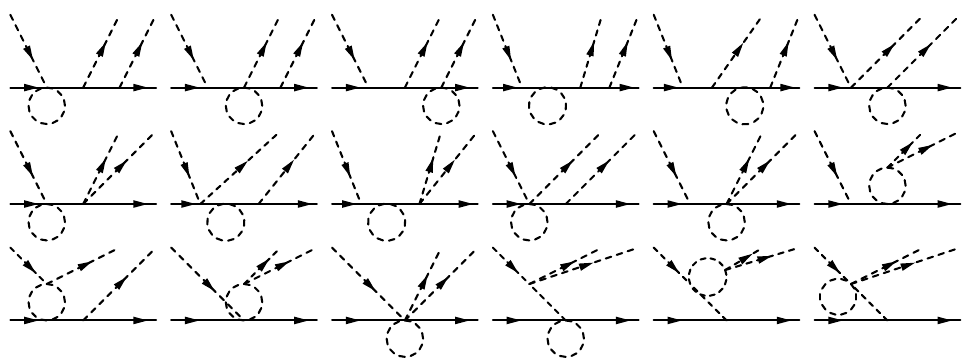}
  \caption{One-loop graphs of the tadpole type contributing to the reaction 
            $\pi N\to\pi\pi N$. For notation see Fig.~\ref{fig:TreeGraphs}.}
  \label{fig:LoopGraphsTadPole}
\end{figure}

\begin{figure}[ht]
  \centering
\includegraphics[width=\textwidth]{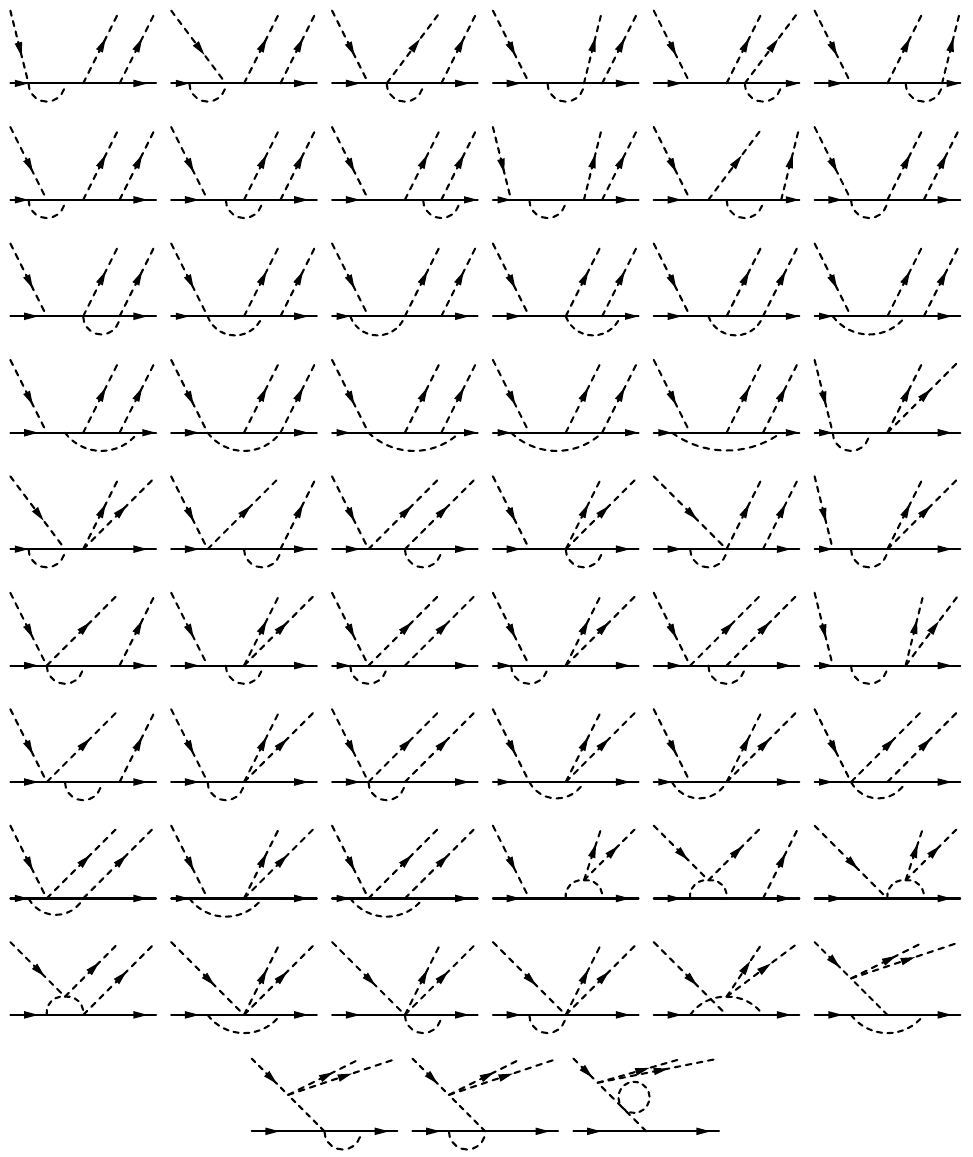}
  \caption{One-loop graphs of the self-energy type contributing to the reaction 
           $\pi N\to\pi\pi N$. For notation see Fig.~\ref{fig:TreeGraphs}.}
  \label{fig:LoopGraphsSelfEnergy}
\end{figure}

\begin{figure}[ht]
  \centering
\includegraphics[width=0.6\textwidth]{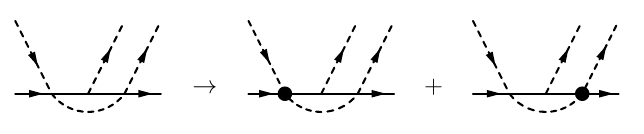}
  \caption{Transition from leading to next-to-leading order loop graphs. 
           For notation see Fig.~\ref{fig:TreeGraphs}.}
  \label{fig:LoopRule}
\end{figure}

\begin{figure}[ht]
  \centering
\includegraphics[width=0.35\textwidth]{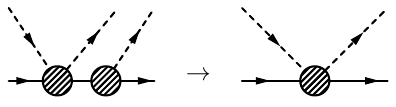}
  \caption{Transition from $\pi N\to\pi\pi N$ graphs to $\pi N\to\pi
    N$ graphs. The shaded blob denotes any possible interaction. 
    For notation see Fig.~\ref{fig:TreeGraphs}.}
  \label{fig:piNRule}
\end{figure}

The relevant tree-level diagrams for the reaction $\pi N\to\pi\pi N$
to order $Q^4$ are shown in
Fig.~\ref{fig:TreeGraphs} while the leading-order loop diagrams at
order $Q^3$ are visualized in Figs.~\ref{fig:LoopGraphsTadPole} and \ref{fig:LoopGraphsSelfEnergy}.
The subleading one-loop diagrams at order $Q^4$ are not shown
explicitly, but can be easily generated by replacing each
leading-order vertex with an even number of pions
from the Lagrangian $\mathcal{L}_{\pi N}^{(1)}$ with a subleading one
from $\mathcal{L}_{\pi N}^{(2)}$ as visualized in
Fig.~\ref{fig:LoopRule}. Notice that there are no vertices with an odd number of
pions in the Lagrangian $\mathcal{L}_{\pi N}^{(2)}$. 
We also do not show here the Feynman diagrams contributing to $\pi
N$-scattering, which can be easily identified by observing that $\pi N\to\pi N$ is a
subprocess of $\pi N\to\pi\pi N$ (see Fig.~\ref{fig:piNRule}), see
also Ref.~\cite{Fettes:2000xg}.

The leading-order tree-level diagrams are constructed solely from
the lowest-order vertices and thus depend only on the well-known LECs
$F_\pi$ and $g_A$. The higher-order tree-level graphs involve insertions of
the LECs $c_i$ from $\mathcal{L}_{\pi N}^{(2)}$, $d_i$ from
$\mathcal{L}_{\pi N}^{(3)}$, $e_i$ from $\mathcal{L}_{\pi N}^{(4)}$
and the purely mesonic LECs $l_i$ from $\mathcal{L}_{\pi\pi}^{(4)}$,
which are known from $\pi\pi$-scattering and other pion observables. 
Specifically, the $\pi N$-scattering amplitudes depend on the
LECs $c_{1,2,3,4}$, $d_{1+2,3,5,14-15}$ and
$e_{14,15,16,17,18}$. These LECs also enter the $\pi N\to \pi\pi N$
amplitudes.  Notice that due to crossing symmetry, 
the contributions proportional to the LECs $e_{14,15,16}$ count
as order-$Q^5$ and for this reason  are set to zero. 
Finally, the $\pi N\to \pi\pi N$ scattering amplitude depends on
additional LECs accompanying the $\pi N$-vertices with 
three pions, namely $d_{4,10,11,12,13,16,18}$ from $\mathcal{L}_{\pi
  N}^{(3)}$ and $e_{10,11,12,13,34}$ from $\mathcal{L}_{\pi N}^{(4)}$.
Note that the LECs $d_4$ and $e_{11,12,13,34}$ only contribute to the channels $\pi^+p\to
\pi^+\pi^0p$ and $\pi^-p\to \pi^0\pi^-p$.
The other LECs contribute to all channels.
Finally, we neglect the contributions proportional to the LEC $e_{35}$,
which appear in the 
amplitudes of both reactions  since the corresponding terms 
actually count as order-$Q^5$.

\section{Fit Procedure}
\label{sec:fitting-procedure}
The amplitudes for the reactions $\pi N\to\pi N$ and 
$\pi N\to\pi\pi N$ depend on several LECs as explained in the previous
section. To extract the LECs $c_i$, $d_i$ and $e_i$ from the data, we
follow the same fit procedure to the available 
pion-nucleon scattering data up to $T<100$~MeV as in
Ref.~\cite{Siemens:2016hdi} but employ two kinds of additional
constraints as discussed below.

\subsection{Constraints from subthreshold parameters}
\label{subsSP}
As a first approach, we consider elastic pion-nucleon scattering but,
differently to our previous study in Ref.~\cite{Siemens:2016hdi}, include in the fitting
procedure additional constraints from the subthreshold region.  
Specifically, we minimize the quantity
\begin{equation}
  \label{eq:31}
  \begin{aligned}
    \chi^2 = \chi^2_{\pi N} + \chi^2_{\mathrm{RS}} \komma
  \end{aligned}
\end{equation}
where $\chi^2_{\pi N}$ is the standard sum of squares
\begin{equation}
  \label{eq:31a}
  \begin{aligned}
    \chi^2_{\pi N} =\sum_i\left( \frac{\mathcal{O}^{exp}_{i}-N_i
      \mathcal{O}^{(n)}_{i}}{\delta \mathcal{O}_i}\right)^2\qquad
  \mathrm{with} \qquad \delta\mathcal{O}_i=\sqrt{(\delta\mathcal{O}^{exp}_i)^2+
(\delta\mathcal{O}^{(n)}_i)^2}\punkt
  \end{aligned}
\end{equation}
The experimental data $\mathcal{O}^{exp}_i$, experimental errors
$\delta\mathcal{O}^{exp}_i$ and normalization factors $N_i$ are
taken from the GWU-SAID data base \cite{Workman:2012hx}.
The quantity $\mathcal{O}^{(n)}_i$ labels the
corresponding observable calculated to chiral order $n$, whereas
the theoretical error $\delta\mathcal{O}^{(n)}_i$ is based on the
truncation of the chiral expansion \cite{Epelbaum:2014efa,Siemens:2016hdi}.
In addition, the quantity $\chi^2_{\mathrm{RS}}$ is defined in
analogy to Eq.~\eqref{eq:31a} as the standard sum of squares, which
includes the eight leading $\pi N$ scattering subthreshold parameters given by the
Roy-Steiner analysis \cite{Hoferichter:2015hva}, namely
$d_{00}^\pm$, $d_{10}^\pm$, $d_{01}^\pm$ and $b_{00}^\pm$.
The Roy-Steiner analysis uses as an input data on the $t$-channel reactions 
(corresponding to the $\pi\pi$, $\bar K K$, $\bar N N$ and other channels),
the $\pi N$ scattering lengths obtained from the analysis of pionic atoms,
the values of the $\pi N$ $S$- and $P$-wave phase shifts at higher energies,
and the values of the $\pi N$  phase shifts for higher partial waves.
Implementing the principles of analyticity and unitarity the $\pi N$ scattering amplitude is continued to the
subthreshold region. Therefore, the subthreshold parameters obtained this way 
contain complementary information to the low-energy $\pi N$ data. 
Among the above-mentioned sources of input information only 
the $\pi N$  phase shifts for higher partial waves
could cause some small amount of double counting of near-threshold data.
The weights in both 
sums of squares in Eq.~\eqref{eq:31} include the experimental error as well as an estimated
theoretical error based on the truncation of the chiral series. The
interested reader is referred to Ref.~\cite{Siemens:2016hdi} for more
details on the fitting procedure. Notice that we choose the values of
the LECs determined by the subthreshold coefficients alone, see
Ref.~\cite{Hoferichter:2015tha}, as a starting point in our iterative
fitting procedure. However, we checked that the final minimum
is independent of the starting point.

One should mention here that adding a theoretical uncertainty in quadrature
as in Eq.~\eqref{eq:31a} is an approximation because of 
correlations of theoretical errors at different data points (see e.g. Refs.~\cite{Furnstahl:2014xsa,Wesolowski:2015fqa}).
We will also make use of the quantity 
\begin{equation}
  \label{eq:31ab}
    \bar\chi^2_{\pi N} =\sum_i\left( \frac{\mathcal{O}^{exp}_{i}-N_i
      \mathcal{O}^{(n)}_{i}}{\delta\mathcal{O}^{exp}_i}\right)^2\,,
\end{equation}
where theoretical errors are not taken into account.

The extracted values of the LECs at orders 
$Q^2$, $Q^3$, $Q^4$ are listed in Table~\ref{tab:FitST} for the
heavy-baryon and covariant schemes along with the corresponding values of the
reduced $\chi_{\pi N}^2$ ($\bar\chi_{\pi N}^2$) with (without) theoretical error.
For the sake of compactness, we restrict ourselves, following Ref.~\cite{Siemens:2016hdi},
to the fits with $T_\pi<100$ MeV that correspond to $1704$ $\pi N$ experimental data points.
The number of degrees of freedom ($\mathrm{dof}$) is equal to the number of data points
minus the number of fitted parameters.
Note that the relative weight of the $\chi^2_{\mathrm{RS}}$ in the total minimal $\chi^2$, e.g. for the fits at order  $Q^4$, does not
exceed $2\%$. Nevertheless, due to small uncertainties of the subthreshold parameters given by the Roy-Steiner analysis
(typically of the order of a few percent),
it's statistical importance is sufficient to influence the fit.
To have a simpler comparison, we also show the values of the LECs
extracted in Ref.~\cite{Siemens:2016hdi} and the corresponding reduced $\chi_{\pi N}^2$( $\bar\chi_{\pi N}^2$).

As can be seen from Table~\ref{tab:FitST}, imposing constraints from subthreshold parameters does
not lead to a qualitative improvement of the statistical uncertainties in the determination of the LECs.
However, strong correlations present in the pure $\pi N$ fit (see
Ref.~\cite{Siemens:2016hdi}) are weakened. 
In a combined fit, no correlation coefficient among the LECs exceeds (by
absolute value) $0.9$. Instead of showing the full
covariance/correlation matrix, we prefer to only discuss the strongly
correlated LECs in the pure $\pi N$ fit (at highest considered order--$Q^4$). In particular, in the HB-NN
counting scheme one observes strong (anti-)correlations between $c_1$ and $c_2$
($0.90$), between $c_2$ and $e_{16}$ ($-0.94$) and between $c_2$ and $d_{1+2}$
($0.94$), which in the fits including the constraints from the
subthreshold region are reduced to ($0.73$), ($-0.62$) and ($0.87$), respectively. In the
HB-$\pi$N scheme, one has a similar situation regarding correlations between
the same set of LECs, which are reduced from ($0.93$), ($-0.93$) and ($0.94$) to
($0.86$), ($-0.59$) and ($0.88$), respectively. In the covariant approach, one
only has a strong correlation between $c_1$ and $c_2$ ($0.92$), which is
reduced to ($0.81$).
 The inclusion of the information about the
subthreshold coefficients in the fits could result in deteriorating
the description of the pion-nucleon scattering data in the physical
region. By comparing the corresponding $\bar\chi_{\pi N}^2$ values 
listed in Table~\ref{tab:FitST} at order $Q^4$, we indeed observe this to be the case in the
HB-$\pi$N approach.\footnote{It is more difficult to interpret the
  results at lower orders due
  to the dependence of the employed theoretical uncertainties on the 
fit results at subsequent chiral orders as explained in detail in
\cite{Siemens:2016hdi}.} 
This can be viewed as an indication that the
HB $\chi$PT fails to provide simultaneous description of the
pion-nucleon scattering amplitude both in the physical and
subthreshold regions which is consistent with the findings of 
Refs.~\cite{Hoferichter:2015tha,Siemens:2016hdi,Siemens:2016jwj}. 
The smallest change in $\bar\chi_{\pi N}^2$ and in the values of the
LECs upon including the information about the subthreshold
coefficients in the fit is observed in the covariant 
approach. This should not come as a surprise given the superior description of 
the subthreshold coefficients based on the LECs determined from 
$\pi N$ scattering data alone in this formulation.

\subsection{Constraints from the reaction $\pi N\to\pi\pi N$}
\label{secPiPiN}
In the second approach, we include additional constraints from the
reaction $\pi N\to\pi\pi N$ such that we minimize
\begin{equation}
  \label{eq:31b}
  \begin{aligned}
    \chi^2 = \chi^2_{\pi N} + \chi^2_{\pi\pi N} +
    \chi^2_{\pi\pi}\komma
  \end{aligned}
\end{equation}
where $\chi^2_{\pi N}$ is defined as in Eq.~\eqref{eq:31a}, $\chi^2_{\pi\pi N}$ is
defined analogously and includes the pion-production total cross section
data up to the maximal energy of $T_\pi<350$ MeV as well as double differential cross
section data at $T_\pi=200$ MeV and $T_\pi=230$ MeV. 
The total cross sections are taken from the compilation \cite{Vereshagin:1995mm} and
from \cite{Kermani:1998gp}, \cite{Lange:1998ti} and
\cite{Prakhov:2004zv}, whereas the double-differential cross sections
with respect to $\Omega_2$ and the pion kinetic energy $T_2=\omega_2 -M_\pi$ 
in the channel $\pi^-p\to\pi^+\pi^-n$ are reported in \cite{Manley:1984zs}. 
The information about $\pi\pi$ scattering data is included indirectly
in $\chi^2_{\pi\pi}$ by using the extracted LECs $l_i$ including
uncertainties as a sum of squares
\begin{equation}
  \chi_{\pi\pi}^2=\sum_i^4\left( \frac{l_i-\bar{l}_i}{\Delta\bar l_i} \right)^2\komma
\end{equation}
 where we used the values for the relevant LECs from $\mathcal{L}_{\pi\pi}^{(4)}$ summarized in \cite{Bijnens:2014lea} \footnote {A recent compilation of  the various results from the lattice simulations can be found in \cite{Aoki:2016frl}.}
\begin{equation}
  \label{eq:46}
  \begin{aligned}
    \bar l_1 &= -0.4\pm 0.6\komma\quad
    \bar l_2 &= 4.3 \pm 0.1\komma\quad
    \bar l_3 &= 2.9 \pm 2.4\komma\quad
    \bar l_4 &= 4.4\pm 0.2\punkt
  \end{aligned}
\end{equation}
Note that $\Delta \bar l_i$ denotes the statistical error such that
we do not employ a theoretical error in $\chi^2_{\pi\pi}$.

As was seen in the analysis of \cite{Siemens:2016hdi}, the $\Delta$
pole at $T_\pi\simeq 190$~MeV and the strong coupling of the $\Delta$ 
to the $\pi N$ sector prevents one from using elastic pion-nucleon
scattering data at energies higher than  $T_\pi\sim 100$~MeV when
extracting the LECs using $\Delta$-less formulations of $\chi$PT.  
The situation in the reaction 
$\pi N\to\pi\pi N$ is somewhat different in the sense that the coupling of the
$\Delta$ to the $\pi\pi N$ sector is very weak as compared to the
coupling to the $\pi N$ sector. This can be seen in the data on 
decay channels of the $\Delta$ \cite{Olive:2016xmw}, where $\Delta\to\pi N$  
contributes to $\sim 100\%$, while the channel $\Delta\to\pi\pi N$ is 
not even listed in Particle Data Group \cite{Olive:2016xmw}. 
Also, the observables such as the total cross sections  do not show any pronounced
structure in the energy region of the $\Delta$ pole.
Notice further that in the reaction $\gamma N\to\pi\pi N$
at threshold one also expects an overwhelming contribution from the
$\Delta$. However, it was shown in Ref.~\cite{Bernard:1994ds} that there are exact cancellations
in the single and double-$\Delta$ tree graphs at threshold that suppress the dangerous
denominator $1/(m_\Delta-m_n-2M_\pi)$.
Thus, it does not appear to be a priori unreasonable to perform fits to $\pi
N\to\pi\pi N$  experimental data in the $\Delta$ region using 
 deltaless  formulations of $\chi$PT. It should, however, be emphasized that
the reaction $\pi N\to\pi\pi N$ has an additional subdecay channel $\Delta\to \pi N$ (via $\pi N\to \pi\Delta$ channel)
for $T_\pi\gtrsim380$~MeV, which might lead to further limitations on
the theory. Moreover, the influence of the Roper resonance may become significant when the 
energy increases. Although its nominal position corresponds to the laboratory
energy of $T_\pi\approx 490$~MeV,
the Roper resonance has a rather large width and a fairly strong coupling to the $\pi\pi N$ channel  \cite{Olive:2016xmw}.
According to the covariant tree-level study in \cite{Jensen:1997em}, the Roper indeed plays a visible 
role in some channels. For other studies of the effects of the $\Delta$ and the Roper resonance
in the considered energy region,
see e.g. Refs.~\cite{Pascalutsa:2002pi,Long:2011rt}.

We performed fits to the discussed $\pi N\to\pi\pi N$ data with incoming pion kinetic energy
$T_{\pi,\pi\pi N}<\{250, 275, 300, 325, 350\}$~MeV, which corresponds
to $\{ 87, 101, 122, 132, 140 \}$ data points, respectively. 
Note that the energy range for calculating $\chi^2_{\pi N}$ ($\bar\chi^2_{\pi N}$) is always taken to be $T_\pi<100$~MeV.
The fitted LECs as functions
of the maximal fitting energy $T_{\pi,\pi\pi N}$ are shown in
Figs.~\ref{fig:LECsQ3}, \ref{fig:LECsQ4P1} and \ref{fig:LECsQ4P2}
while the reduced $\chi_{\pi N}^2$ ($\bar\chi_{\pi N}^2$) and
$\chi_{\pi\pi N}^2$ ($\bar\chi_{\pi\pi N}^2$) with (without)
theoretical errors as a function of $T_{\pi,\pi\pi N}$ is plotted in
Fig.~\ref{fig:RedChiSq}. Here, the number of degrees of freedom (dof) for the $\pi N\to\pi\pi N$
reaction is defined as the number of the data points for this reaction minus the
number of additional parameters not appearing in the $\pi N$ scattering amplitude.
We interpret the stability of the fit against the maximum fitted energy 
as an indicator of convergence of the chiral expansion
in the considered energy region and of the correct choice of 
the breakdown scale $\Lambda_b$ (a similar strategy was introduced in Ref.~\cite{Wesolowski:2015fqa}). We do not employ here
more sophisticated methods based on the Bayesian approach
as it was done e.g. in Refs.~\cite{Furnstahl:2015rha,Melendez:2017phj}. 
While the fits at
$Q^3$ exhibit a plateau-like behaviour of the extracted LECs as well
as of
the $\chi_{\pi N}^2/$dof and $\chi_{\pi\pi N }^2/$dof with regard to
the maximal energy of the $\pi N\to\pi\pi N$ data,
the $\chi_{\pi\pi N}^2/$dof and the extracted LECs at $Q^4$
deviate rather strongly from a constant behaviour when the energy is
increased. Optimistically, only the fit results up to 275~MeV may be
regarded as reasonably stable. Moreover, as shown in the lowest row of
Fig.~\ref{fig:RedChiSq}, the description of the $\pi N\to\pi\pi N$ data actually
deteriorates at order $Q^4$ as compared to the order $Q^3$ except for 
the results within the covariant approach at energies below
$300~$MeV. The problem can be traced back to the 
large values of some of the $d_i$, which are preferred by
the $\pi N$ scattering data at order $Q^4$ and seem to be  
in conflict with the $\pi N\to\pi\pi N$ data. This especially applies
to the linear combination $d_{14-15}$, which 
changes its value from $d_{14-15} \sim -6~$GeV$^{-2}$ at $Q^3$ to 
$d_{14-15} \sim -10~$GeV$^{-2}$ at $Q^4$ in the covariant approach. 
We, however, found that the magnitude of the linear combination $d_{14-15}$ at $Q^4$ has to be
much smaller in order to improve the 
convergence pattern of the chiral expansion in the single-pion production.
Notice that the low-energy constants contributing to elastic pion-nucleon
scattering are known to become significantly smaller in magnitude upon
explicit treatment of the  $\Delta$-resonance. This effect of resonance saturation was observed, in particular,
in Ref.~\cite{Siemens:2016hdi}, where the leading-order
$\Delta$-contributions have been included. Unfortunately, as will be discussed in section~\ref{sec:LECs}, the
analogous simplified inclusion of the $\Delta$-resonance in the $\pi
N\to\pi\pi N$ reaction is less straightforward due to the appearance of
a number of additional free parameters. Moreover, as already mentioned
above, one cannot a priori exclude the possibility that the Roper-resonance provides significant 
contributions to some of the $3\pi NN$ LECs as well, while its contribution to the leading 
$2\pi NN$ LECs  $c_{1,2,3,4}$ is known to be marginal~\cite{Bernard:1996gq}.
A consistent inclusion of the $\Delta$ and Roper resonances in the
framework of $\chi$PT, which may be needed to increase the
applicability range of the theory, is, however,  beyond the scope of this paper.

The values of the LECs extracted at orders 
$Q^2$, $Q^3$, $Q^4$ are collected in Tables~\ref{tab:FitpipiN1} and \ref{tab:FitpipiN2} for all
considered approaches along with the corresponding values of the
reduced $\chi_{\pi N}^2$ and $\chi_{\pi\pi N}^2$.
To demonstrate the impact of the constraints from the
reaction $\pi N\to\pi\pi N$, we restrict ourselves to the fits with
$T_{\pi,\pi\pi N}<275$~MeV where our results are fairly stable. 
 
In general, the change of the LECs as compared to the pure $\pi N$ fit
appears to be small. This can be traced back to the almost complete
decoupling of the $\pi N\to\pi\pi N$ component of the $\chi^2$ from 
the $\pi N \to \pi N$ one caused by  the large theoretical uncertainties 
in the $\pi N\to\pi\pi N$ sector. Also the statistical errors and correlations of the
LECs remain almost unchanged. In addition, we observe strong 
anticorrelations between the LECs $d_{10}$, $d_{12}$ and
$d_{11}$,$d_{13}$, see Table \ref{tab:Q4pipiNLECsCorr} for the results
in the covariant approach.

\section{Naturalness of the LECs}
\label{sec:LECs}
Let us comment on 
the extracted numerical values of the 3$\pi$NN LECs given in
Table~\ref{tab:FitpipiN2}. Indeed at first sight they appear to be rather large if we would be using the very 
naive estimation based on the naturalness assumption,
\begin{equation}
  \label{eq:8}
c_i \sim \frac{1}{\Lambda_b} \sim 2\,\mathrm{GeV}^{-1}\komma\qquad  d_i  \sim \frac{1}{\Lambda_b^2} \sim 3\,\mathrm{GeV}^{-2}
\komma\qquad  e_i \sim \frac{1}{\Lambda_b^3} \sim 5\,\mathrm{GeV}^{-3} \,,
\end{equation}
where $\Lambda_b = 600$ MeV is used to estimate the breakdown scale of
the chiral expansion. For the unnaturally large 2$\pi$NN LECs, the origin of their enhancement 
can be traced back to the implicit treatment of the $\Delta$
resonance~\cite{Bernard:1996gq}.
As shown in \cite{Krebs:2007rh,Siemens:2016hdi}, the explicit inclusion of
the  leading $\Delta$-pole diagrams leads to natural values for all LECs. Following the same
strategy, we have repeated the fits including the leading
$\Delta$-pole diagrams in the reaction 
$\pi N \to\pi \pi N$ while setting the additional LECs from the
$\Delta$ sector to their large-$N_c$ values, namely $h_A=1.35$ and $g_1=2.29$.
The results for the 3$\pi$NN LECs at order $Q^4 + \delta^1$ are given in Table
\ref{tab:Q4pipiNLECsCorr}, whereas the 2$\pi$NN LECs are not given
explicitly but are in very good agreement with the ones determined in \cite{Siemens:2016hdi}.
As can be seen from the table, the LECs $d_{i}$ still remain large
whereas the LECs $e_i$ do indeed become more natural as
compared to the deltaless fits.  Notice further that the statistical errors
and the correlations among the LECs $d_{10,11,12,13}$ get enhanced
(see Table~\ref{tab:Q4pipiNLECsCorr})
upon including the $\delta^1$-contributions.

To get further insights into the observed pattern,  it
is instructive to 
consider the NLO contributions of the $\Delta$ and Roper resonances to the
relevant LECs, which are explicitly given in appendix \ref{satur}. 
These expressions are based on the effective Lagrangian
\begin{equation}
\label{eq:7}
  \begin{aligned}
\mathcal{L}_{\pi\Delta}^{(1)} &=- \bar{\Psi}_i^\mu\Big[
    (\i\slashed{D}^{ij}-{m_\Delta} \delta^{ij} )g_{\mu\nu}-\i (\gamma_\mu
    D_\nu^{ij}+\gamma_\nu D_\nu^{ij})+\i \gamma_\mu
    \slashed{D}^{ij}\gamma_\nu+{m_\Delta} \delta^{ij}\gamma_\mu \gamma_\nu \\
    &\qquad\qquad+\frac{{g_1}}{2}g_{\mu\nu} \slashed{u}^{ij} \gamma_5 
    \Big]\Psi^\nu_j\komma\\
    \mathcal{L}_{\pi N \Delta}^{(1)}&= h_A\,\bar{\Psi}_\mu^i\Theta^{\mu\alpha}(z_0)w_\alpha^i\Psi
    +\mathrm{h.c.}\komma\\
    \mathcal{L}_{\pi N \Delta}^{(2)}&= \bar{\Psi}_\mu^i \Theta^{\mu
      \alpha}(z_1)\Big[ 
      \frac{b_4}{2}w_\alpha^i w_\beta^j \gamma^\beta\gamma_5\tau^j +
      \frac{b_5}{2}w_\alpha^j w_\beta^i \gamma^\beta\gamma_5\tau^j \Big]\Psi
    +\mathrm{h.c.}\komma\\
    \mathcal{L}_{\pi R}^{(1)}&=\bar{\Psi}_R \left[ \i
      \slashed{D}-m_R+\frac{g_{RR}}{2}\slashed{u}\gamma_5 \right]\Psi_R\komma\\
    \mathcal{L}_{\pi R}^{(2)}&= \bar{\Psi}_R \left[c^R_1\Braket{\chi_+} +
      \frac{c^R_2}{8m^2} (-\braket{u_\mu
        u_\nu}D^{\mu\nu}+\mathrm{h.c.} ) +
      \frac{c^R_3}{2}\Braket{u\cdot u}-\frac{c^R_4}{2} \sigma^{\mu\nu}[u_\mu, u_\nu]\right]
    \Psi_R\komma\\
    \mathcal{L}_{\pi RN}^{(1)}&=\bar{\Psi}_R \left[\frac{g_{RN}}{2}\slashed{u}\gamma_5 \right]\Psi +\mathrm{h.c.}\komma\\
    \mathcal{L}_{\pi R \Delta}^{(1)}&=
 g_{R\Delta}\,\bar{\Psi}_\mu^i\Theta^{\mu\alpha}(z_2)w_\alpha^i\Psi_R
    +\mathrm{h.c.}\komma
  \end{aligned}
\end{equation}
where the Roper contributions are introduced in a close analogy
with the pion-nucleon Lagrangian in Ref.~\cite{Fettes:2000gb}, as first done in  \cite{Borasoy:2006fk} ,
and the pion-nucleon-$\Delta$ Lagrangian is taken from Refs.~\cite{Hemmert:1997wz,Hemmert:1997ye}.
Details on the notation used in Eq. \eqref{eq:7} can be found in Refs.~\cite{Siemens:2014pma,
Fettes:2000gb,Hemmert:1997wz,Hemmert:1997ye}. Note that we set the off-shell
parameters $z_i=0$ in the
explicit expressions.
The numerical contributions of the 
$\Delta$ and Roper resonances to the considered LECs are summarized 
in Table~\ref{tab:Q4SatFlow}. The numerical values are obtained 
by assuming natural values for the unknown LECs entering
these expressions. In the $\Delta$-sector, we fix $h_A=1.35$ and
$g_1=\pm 2.29$ to their large $N_C$ values and employ $b_4=b_5=\pm 1$.
In the Roper sector, we fix $g_{RN}=0.35$ as determined by the decay
width of $R\to\pi N$ \cite{Borasoy:2006fk} and assume
$g_{RR}=g_{R\Delta}=c_i^R=\pm1$.
As can be seen, the contributions to the LECs from the leading-order $\Delta$-pole
diagrams employed in our fits ($g_1=2.29$) are relatively small for
the large LECs $d_{10,11,12,13}$ while quite substantial for the
large LECs $e_{10,11,12,13}$. This pattern is consistent with the 
$Q^4 + \delta^1$ values of the LECs listed in Table
\ref{tab:Q4pipiNLECsCorr}. Concerning the higher-order
contributions, we find some potentially large terms proportional
to $b_4$, $b_5$ as well as to $g_{R\Delta}$. 
The remaining contributions of the Roper resonance are
rather small and can be neglected. Note that the LECs $c_i$ were
redefined to absorb redundant contributions proportional to certain
linear combinations of 
$e_i$ \cite{Siemens:2016hdi}, which induces the explicit
$\mu$-dependence of $c_i$ even in the HB approach. 

Having established that the large values of the 3$\pi$NN LECs
$d_i$ cannot be explained by means of resonance saturation, it is instructive 
to address their sensitivity to the choice of the renormalization
scale $\mu$. 
To be specific, we consider the changes
in the values of the LECs by changing the renormalization scale $\mu$
from $M_\pi$ to $m_N$,
\begin{equation}
  \label{eq:9}
  \Delta x \equiv x \big|_{\mu=m_N}-x \big|_{\mu=M_\pi}\komma \qquad
  \overline{\Delta x} = \frac {x(\mu=M_\pi)}{\Delta x} \,,
\end{equation}
where $x\in\{c_i, d_i,e_i\}$. The quantity ${\Delta x}$ gives the
absolute  change of a LEC $x$, whereas $\overline{\Delta x}$
is a measure of its relative change.
Notice that throughout this work, we follow the  convention by
choosing $\mu=M_\pi$. The  renormalization-group (RG) flow of the LECs 
is determined by the corresponding dimensionless
$\beta$-functions. At one-loop level, one finds 
 \begin{equation}
  \label{eq:10}
 \Delta x =  \frac{\beta_x}{32\pi^2 F_\pi^2}\log\Big(\frac{M_\pi^2}{m_N^2}\Big)\,,
\end{equation}
and the $\beta$-functions can be found in \cite{Siemens:2016hdi} for both
the covariant and heavy-baryon approaches. As can be seen from Table~\ref{tab:Q4SatFlow}, 
the shifts in the 3$\pi$NN LECs under the
considered change of the renormalization point appear to be much larger than
the ones in the 2$\pi$NN LECs and are, in most cases, of the same
order of magnitude as the LECs themselves. 
This provides yet another indication that the observed large size of these LECs is not related 
to the implicit treatment of the $\Delta$ and Roper
resonances but is rather caused by the corresponding dimensionless
$\beta$-functions being numerically large. While such enhancement of the $\beta$-functions
may emerge due to combinatorial reasons such as the products of spin and/or isospin matrices or powers of $g_A$, 
which could affect the convergence pattern of the chiral expansion, 
it could also come from  the adopted form of the effective Lagrangian which is a matter of convention. Thus, one
cannot a priori exclude the possibility that the large values of the
LECs simply reflect the convention employed in the effective
Lagrangian. More precisely, the vertices with many pions contain factorial factors
that are not reflected in the corresponding terms in the effective Lagrangian.
Another interesting observation is that the LECs $c_i$
decrease in magnitude when the renormalization scale is increased, while
the LECs $d_i$ show the opposite behaviour and grow in magnitude 
when increasing the renormalization scale. For the LECs $e_i$ one has a
mixed pattern, where the 2$\pi$NN LECs increase and the 3$\pi$NN LECs
decrease in magnitude.

We now further elaborate on the possibility that the large numerical
values of the 3$\pi$NN LECs are caused by the convention employed in
the effective Lagrangian as explained before.
Due to the complexity of the $\pi N\to\pi\pi N$ amplitudes involving
several energy scales, it is, however, difficult to estimate the contributions
from each individual LEC and to identify possible numerical enhancements of
this sort. One simple approach is to perform an expansion 
around the threshold point $\omega_2= \omega_3=M_\pi$ and 
$\vec{q_1}\cdot\vec{q_2}=\vec{q_1}\cdot\vec{q_3}=\vec{q_2}\cdot\vec{q_3}=0$,
such that each Taylor coefficient of that series only involves the
scales $M_\pi$ and $m_N$. In the following we will consider one
representative example. A threshold expansion of the HB-NN amplitude for 
the channel $\pi^+p\to\pi^+\pi^-n$ denoted by III gives
\begin{equation}
  \label{eq:2}
  \begin{aligned}
    T_{III}\simeq&\Big[ - \frac{4 \i}{F_\pi^3}\Big(
    2(d_{10}+d_{12})+d_{11}+d_{13}
    +6g_A(d_{1+2}+d_3+d_5)) \Big) M_\pi^2 +\\
    &+ \frac{2\i}{F_\pi^3} \Big(
    2 d_{10}+d_{11}+3g_A(2d_{1+2}-d_{14-15})\Big) \vec{q_1}\cdot
    \vec{q_3}+\dots\Big] S\cdot q_2+\dots\komma
  \end{aligned}
\end{equation}
where we only display contributions proportional to the $d_i$. As can be
seen, the  contribution from the 3$\pi$NN LECs, namely $d_{10}$,
$d_{11}$, $d_{12}$ and $d_{13}$, is suppressed by large numerical
factors relative to  the one of the 2$\pi$NN LECs. 
This is an indication that the large numerical values of the
3$\pi$NN LECs resulting from our fits merely reflect 
the chosen normalization
in the effective Lagrangian. Also, one should keep in mind that only 
combinations of these LECs
show up. Given that there are some non-negligible correlations between these LECs, looking at the
individual values might be misleading. Similar 
observations can be made for the 3$\pi$NN LECs $e_i$. All this requires
more detailed studies that go beyond the scope of this work. However, 
we would like to stress that especially in  the nucleon sector where
multiple powers of the axial coupling constant enter, the usage of the very naive assumption about the natural size of the 
LECs, Eq.~(\ref{eq:8})
when increasing the order one is working with, should be taken with a grain of salt.

\section{Predictions}
\label{sec:predictions}
Based on the LECs extracted in the previous section, we are now
in the position to make predictions for various observables. 
In particular, we focus on the threshold and subthreshold $\pi N$  coefficients.
The relation of the $\pi N$ amplitude to the subthreshold parameters 
is given in section \ref{sec:SubPar}. 
The threshold expansion of the amplitudes 
\begin{equation}
  \label{eq:3thr}
\mathrm{Re}\, D^\pm=\sum^{\infty}_{n,m=0} D^\pm_{mn} \vec q^{2m}t^n\komma\quad
\mathrm{Re}\, B^\pm=\sum^{\infty}_{n,m=0} B^\pm_{mn} \vec q^{2m}t^n
\end{equation}
is related to the threshold parameters via the expansion of the partial wave amplitude
\begin{equation}
  \label{eq:5}
  \mathrm{Re}\, T_{l\pm}= q^{2l+1}(a_{l\pm}+b_{l\pm} \vec q^2+\dots)
\end{equation}
such that the parameters of interest are given by
\begin{equation}
  \label{eq:6}
  \begin{aligned}
    a^\pm_{0+}&=\frac{D^\pm_{00}}{4\pi (1+\alpha)}\komma \qquad
    b^\pm_{0+}=-\frac{(2-\alpha)D^\pm_{00}+8D^\pm_{01}m_N^2\alpha-4D_{10}m_N^2\alpha-2B_{00}m_N\alpha^2}{16\pi m_n^2\alpha(1+\alpha)}\komma\\
    a^\pm_{1+}&=-\frac{B^\pm_{00}-4D^\pm_{01}m_N}{24\pi m_N
      (1+\alpha)} \komma\qquad
    a^\pm_{1-}=-\frac{3D^\pm_{00}-8D^\pm_{01}m_N^2-B^\pm_{00}m_N(4+6\alpha)}{48\pi
      m_N^2 (1+\alpha)} \komma
  \end{aligned}
\end{equation}
with $\alpha=M_\pi/m_N$.

Our  results for the sub- and threshold parameters based on  
the different fit
approaches are collected in Table \ref{tab:SubThrPara} and
\ref{tab:ThrPara}, respectively.
As one would expect, the description of the subthreshold parameters improves when using them as an additional
constraint and remains similar in quality to the pure $\pi N$ fit when performing a combined fit with $\pi N\to\pi\pi N$ reaction.
In general, the agreement with the subthreshold and threshold parameters obtained from the Roy-Steiner (RS) equations is 
better in the covariant approach. This scheme also yields results which are more stable against introducing additional constraints
as compared with the HB $\chi$PT formulations.

Next, our predictions for the $\pi N$ phase shifts in $S$ and $P$
partial waves up to pion energies of $100$~MeV are given
in Figs.~\ref{fig:SwavesST} and \ref{fig:Swaves} for the two different
fit strategies in comparison with the RS results of
Ref.~\cite{Hoferichter:2015hva}. A comparison of Fig.~\ref{fig:SwavesST}
and Fig.~9 of Ref.~\cite{Siemens:2016hdi} reveals that the
additional constraints from the subthreshold coefficients have little
impact on the phase shifts in the physical region when using the
covariant $\chi$PT formulation, while the changes are more visible in
the two considered HB approaches. These observations are  in line with
the conclusions of section~\ref{subsSP}. Further, as already pointed
out above, using the additional constraints from the reaction $\pi N
\to \pi \pi N$ has almost no effect on the description of $\pi N$
scattering in the physical region within the employed fitting procedure.
As a consequence, the predictions for the phase shifts in
Fig.~\ref{fig:Swaves} are almost identical to the ones shown in Fig.~9
of Ref.~\cite{Siemens:2016hdi} for all considered counting schemes. 

We now turn to the reaction $\pi N\to\pi\pi N$. As explained in
section~\ref{secPiPiN}, we are unable to obtain simultaneously  a good description
of both the $\pi N \to \pi N$ and $\pi N \to \pi \pi N$ data at order
$Q^4$, which is mainly due to the large values of some of
the  2$\pi$NN LECs $d_i$ preferred by the elastic scattering data
being seemingly incompatible with the single pion production data. 
Here and in what follows, we, therefore, show only a few
representative examples for observables. Our results for the total
cross section in five channels are shown in Fig.~\ref{fig:sigmatot}. 
While the description of the data at low energies used in the fit is
fairly good, one observes a strong overestimation of the cross section
at higher energies, which is particularly pronounced in the $\pi^+ p
\to \pi^+ \pi^+ n$ and $\pi^- p \to \pi^0 \pi^- p$ channels. 
While the covariant approach shows the smallest deviations from the
data, one observes no improvement (at both orders $Q^3$ and $Q^4$)  as compared with the tree-level
calculations of Ref.~\cite{Siemens:2014pma}. In Figs.~\ref{fig:sigmadiffW1}-\ref{fig:sigmadiffW4}, we also
show selected observables in the channel $\pi^- p
\to \pi^+ \pi^+ n$ which may be viewed as representative
examples. Specifically, the angular
correlation function $W$ is shown as a function of the final dipion mass
squared $M_{\pi \pi}^2$ for fixed angles $\theta_1$ and $\theta_2$
($\theta_1$ and $\phi_2$) in Figs.~\ref{fig:sigmadiffW1} and \ref{fig:sigmadiffW2} 
(Figs.~\ref{fig:sigmadiffW3} and \ref{fig:sigmadiffW4}) in comparison
with the  data from 
Ref.~\cite{Muller:1993pb}. Further, our
predictions for the single-differential cross sections with respect to
$M_{\pi \pi}^2$ and $t$ are plotted in
Figs.~\ref{fig:sigmadiffdMsq2} and \ref{fig:sigmadiffdt2} in
comparison with the data from
 Ref.~\cite{Kermani:1998gp}.  We refer the reader to
 Ref.~\cite{Siemens:2014pma} for details on the kinematics and for the
 definitions of various observables in this reaction. Comparing our
 predictions with the tree-level
 calculations reported in Ref.~\cite{Siemens:2014pma}, we observe a
 clear improvement for the angular correlation at $\theta_1 =
 76^\circ$
and $\theta_2 = 66.7^\circ$,   $\theta_2 = 39.7^\circ$ as well as 
at $\theta_1 =
 71^\circ$
and $\theta_2 = 69.4^\circ$,   $\theta_2 = 41.5^\circ$, see the lower
two panels of Figs.~\ref{fig:sigmadiffW1} and \ref{fig:sigmadiffW2}. 
In all remaining cases shown in
Figs.~\ref{fig:sigmadiffW1}-\ref{fig:sigmadiffdt2}, 
the description of the data appears to be comparable to
the one reported in  Ref.~\cite{Siemens:2014pma}. 

As already mentioned in section~\ref{sec:fitting-procedure}, the 
most probable reason for a slower convergence of the chiral expansion
at higher energies are the missing contributions of the $\Delta$ and 
Roper resonances.
A full-fledged inclusion of the $\Delta$ and Roper resonances would 
require calculating a number of tree-level and loop diagrams and
adjusting many additional parameters, which goes beyond the scope of
this work.  Instead, we perform here a simplified partial inclusion of the
$\Delta$ resonance by taking into account
the leading $\Delta$-pole diagrams in the $\pi N$ elastic channel
(as was done in Ref.~\cite{Siemens:2016hdi}) and
in the $\pi N\to\pi\pi N$. 
To avoid the introduction of additional parameters, we set the 
constants $h_A$ and $g_1$  to their large-$N_c$ values, see section~\ref{sec:LECs}.
Note that although the sign of $g_1$ can be fixed by large-$N_c$
constraints, we also checked that using the opposite sign, $g_1=-2.29$,
has no substantial effects
on the results because the $\Delta$-pole contribution to the $\pi
N\to\pi\pi N$ amplitude appears to be rather small, consistent with
the findings in Ref.~\cite{Bernard:1994ds} for the reaction $\gamma N\to \pi\pi N$.
On the other hand, the inclusion of the leading $\Delta$-pole diagrams
in the $\pi N\to\pi N$ channel influences indirectly the results in the 
$\pi N\to\pi\pi N$ channel since
the obtained LECs (in particular $d_i$) become  smaller in line with the
resonance saturation, see the discussion in section~\ref{sec:LECs}.
As a result, the description of the $\pi N\to\pi\pi N$ data improves
significantly.  This is illustrated with the example of the total
cross sections for all five channels in Fig.~\ref{fig:sigmatotDelta}. 
The $\chi^2$ ($\bar\chi^2$) also show a dramatic improvement for both
$\pi N \to \pi N$ and $\pi N\to\pi\pi N$ reactions, and their dependence on a
maximum energy in the $\pi\pi N$ channel 
becomes much more flat (cf. Fig.~\ref{fig:RedChiSqWithDelta}). 
This indicates a potentially better convergence  
of the chiral expansion in the presence of explicit $\Delta$ degrees of freedom.
The fact that the $\bar\chi^2_{\pi\pi N}$ slightly increases at higher
energy could signal the importance of the Roper resonance, 
which we do not take into account explicitly.

We have reached the values of reduced $\bar\chi^2$ around $2(3)$
for the $\bar\chi_{\pi N}^2/$dof ($\bar\chi_{\pi \pi N}^2/$dof) at order $Q^4+\delta^1$.
To obtain the values of reduced $\bar\chi^2$ even closer to $1$ it would likely be
necessary to refine the existing data base by rejecting inconsistent data sets
as it is commonly done in the nucleon-nucleon sector \cite{Stoks:1993tb,Perez:2013jpa}. 
Consider a more established case of the $\pi N$ data. In this case
our $\bar\chi^2_{\pi N}$ does not differ much from the $\chi^2_{\pi N}$,
which indicates that the errors are dominated by the experimental uncertainties.
On the other hand, we utilize the GWU-SAID data base 
\cite{Workman:2012hx} without change, and the reduced $\chi^2$ of the GWU-SAID partial-wave analysis \cite{Arndt:2006bf}
(for the same data base and the energy region relevant for our study) is equal to $\chi^2/\mathrm{dof}\approx 1.8$.
Clearly, we will not be able to obtain  $\chi^2/\mathrm{dof}$ smaller than this value
without modifying the data base by means of rejecting some inconsistent data.
This goes, however, beyond the scope of the present work. 

\section{Summary and outlook}
\label{sec:sum}

The main results of our paper can be summarized as follows:
\begin{itemize}
\item
We have extended our analysis of  pion-nucleon scattering in chiral perturbation 
theory to the  full one-loop order ($Q^3$ and $Q^4$) reported in 
Ref.~\cite{Siemens:2016hdi} by imposing additional 
constraints from the subthreshold parameters calculated by means of
the Roy-Steiner equations in Ref.~\cite{Hoferichter:2015hva} and from the  combined fit 
with the $\pi N\to\pi\pi N$ reaction at low energies. We have
considered all three formulations of  $\chi$PT , namely the
heavy-baryon schemes HB-NN, HB-$\pi$N and the
covariant version. For the first time, the $\pi N\to\pi\pi N$ scattering
amplitude has been calculated at the chiral order $Q^4$. 
The fits to the combined data sets were performed employing the  
approach for estimating the theoretical
uncertainty from the truncation of the chiral expansion introduced in
Ref.~\cite{Epelbaum:2014efa}.
\item
For the combined fit with the Roy-Steiner subthreshold parameters the
extracted low-energy constants are found to have similar statistical uncertainties 
as in the fit to  $\pi N$ scattering data alone. However, we found
that taking into account the additional constraints in the
subthreshold region allows to strongly suppress the amount of
correlations between some of the LECs. 
The description of the subthreshold parameters in the combined fit  is 
obviously improved whereas 
the $\pi N$ data in the physical region are reproduced slightly
worse. The smallest change in the $\bar\chi^2$ (without theoretical
errors) and in  the
values of LECs is observed for the covariant formulation of $\chi$PT. 
\item
For the combined fit with  the $\pi N\to\pi\pi N$ reaction, the extracted 
low-energy constants
already contributing to the elastic $\pi N$ amplitude and  their
statistical uncertainties remain nearly unchanged.
As in the case of the constraints from the subthreshold region, strong
correlations among LECs are found to be reduced.
Some of the new LECs that give contributions to the $\pi N\to \pi\pi
N$ amplitude appear to be ``unnaturally'' large in magnitude.
However, our analysis shows that the corresponding LECs appear in the
scattering amplitudes in linear combinations, which are suppressed by
large numerical factors as compared  to the other LECs. As a result,
we do not observe any unnatural enhancement of their contributions to
the scattering observables.  
\item
Using the results of the combined fit to the $\pi N\to\pi N$ and $\pi
N\to\pi\pi N$ reactions,  
we confront the results of our calculations with the experimental data
for various $\pi N\to\pi\pi N$ observables 
as well as for the $\pi N$ phase shifts. For all three formulations of $\chi$PT, 
we obtain a satisfactory description of the experimental/empirical data and 
a reasonable convergence pattern. The agreement with the
data becomes worse as the energy rises. This most probably indicates the importance of the
$\pi\Delta$ channel and the Roper pole, which we do not take  into account explicitly.
A simplified, partial inclusion of the $\Delta$ resonance via tree-level pole diagrams 
leads to a significant improvement
in the description of the data in both $\pi N\to\pi N$ and $\pi N\to\pi\pi N$ 
channels in accordance with this assumption.
We anticipate that a rigorous treatment of the $\Delta$ and Roper 
resonances as explicit degrees of freedom within $\chi$PT, extending the tree-level
study of Ref.~\cite{Jensen:1997em},
will improve convergence of the theory and agreement with the data for 
two considered reactions
and will make it possible to extend the energy region of applicability
of chiral perturbation theory, see also Ref.~\cite{Siemens:2016jwj}. 
Work along these lines is in progress.

\end{itemize}

\section*{Acknowledgments}
This work was supported in part by the Heisenberg-Landau program, the DFG and NSFC through funds provided to the
Sino-German CRC 110 ``Symmetries and the Emergence of Structure in QCD" (NSFC
Grant No.~11621131001, DFG Grant No.~TRR110), 
the ERC project 259218 NUCLEAREFT, the Ruhr University Research
School PLUS, funded by Germany's Excellence Initiative [DFG GSC 98/3],
by the Chinese Academy of Sciences (CAS)
President's International Fellowship Initiative (PIFI) (Grant No.
2017VMA0025) and by the National Science Foundation
under Grant No.~NSF PHY11-25915.sizeable

\appendix
\section{Resonance saturation of the LECs}
\label{satur}
Below, we give the explicit expressions for resonance contributions to the
various LECs. 
The contributions of the $\Delta$ degrees of freedom to the
$\pi$N LECs read:
\begin{equation}
  \label{eq:11}
  \begin{aligned}
    c_{1,\Delta}&=0\,, \\
    c_{2,\Delta}&=\frac{4 h_A^2 m_N^2}{9 (m_N-m_\Delta) m_\Delta^2}\,, \\
    c_{3,\Delta}&=-\frac{4 h_A^2}{9 (m_N-m_\Delta)}\,, \\
    c_{4,\Delta}&=\frac{2 h_A^2}{9 (m_N-m_\Delta)}\,, \\
    d_{1+2,\Delta}&=-\frac{h_A^2 \left(2 m_N^2-3 m_N m_\Delta+3 m_\Delta^2\right)}{18 (m_N-m_\Delta)^2 m_\Delta^2}\,, \\
d_{3,\Delta}&=\frac{h_A^2 m_N^2}{9 (m_N-m_\Delta)^2 m_\Delta^2}\,, \\
d_{4,\Delta}&=-\frac{g_A h_A^2}{36 (m_N-m_\Delta)^2}-\frac{5 g_1 h_A^2 \left(m_N^2-2 m_N m_\Delta-4 m_\Delta^2\right)}{324 (m_N-m_\Delta)^2 m_\Delta^2}\\&-\frac{h_A b_{4}}{18 (m_N-m_\Delta)}+\frac{h_A b_{5}}{18 (m_N-m_\Delta)}\,, \\
d_{5,\Delta}&=-\frac{h_A^2 (2 m_N+m_\Delta)}{36 (m_N-m_\Delta) m_\Delta^2}\,, \\
d_{10,\Delta}&=-\frac{g_A h_A^2}{3 (m_N-m_\Delta)^2}+\frac{g_1 h_A^2 \left(m_N^2-2 m_N m_\Delta+4 m_\Delta^2\right)}{27 (m_N-m_\Delta)^2 m_\Delta^2}\\&-\frac{2 h_A b_{4}}{3 (m_N-m_\Delta)}-\frac{5 h_A b_{5}}{9 (m_N-m_\Delta)}\,, \\
d_{11,\Delta}&=\frac{h_A b_{5}}{9 m_N-9 m_\Delta}-\frac{g_1 h_A^2 \left(4 m_N^2-8 m_N m_\Delta+11 m_\Delta^2\right)}{81 (m_N-m_\Delta)^2 m_\Delta^2}\\&+\frac{g_A h_A^2}{9 (m_N-m_\Delta)^2}+\frac{2 h_A b_{4}}{9 (m_N-m_\Delta)}\,, \\
d_{12,\Delta}&=\frac{g_A h_A^2 m_N (2 m_N+m_\Delta)}{9 (m_N-m_\Delta)^2 m_\Delta^2}+\frac{g_1 h_A^2 m_N^2 \left(8 m_N^2+2 m_N m_\Delta-19 m_\Delta^2\right)}{81 (m_N-m_\Delta)^2 m_\Delta^4}\\&+\frac{2 h_A b_{4} m_N (2 m_N+m_\Delta)}{9 (m_N-m_\Delta) m_\Delta^2}+\frac{h_A b_{5} m_N (2 m_N+3 m_\Delta)}{9 (m_N-m_\Delta) m_\Delta^2}\,, \\
d_{13,\Delta}&=\frac{h_A b_{5} m_N (2 m_N-3 m_\Delta)}{9 (m_N-m_\Delta) m_\Delta^2}-\frac{g_1 h_A^2 m_N^2 \left(4 m_N^2+6 m_N m_\Delta-17 m_\Delta^2\right)}{81 (m_N-m_\Delta)^2 m_\Delta^4}\\&-\frac{2 h_A b_{4} m_N}{9 m_N m_\Delta-9 m_\Delta^2}-\frac{g_A h_A^2 m_N}{9 (m_N-m_\Delta)^2 m_\Delta}\,, \\ 
d_{14-15,\Delta}&=\frac{2 h_A^2 m_N}{9 (m_N-m_\Delta)^2 m_\Delta}\\
d_{16,\Delta}&=0\,, \\
\end{aligned}
\end{equation}
\begin{equation}
\begin{aligned}
e_{10,\Delta}&=\frac{g_A h_A^2 (3 m_N-m_\Delta)}{72 m_\Delta (-m_N+m_\Delta)^3}+\frac{5 g_1 h_A^2 m_N \left(2 m_N^2-9 m_N m_\Delta+17 m_\Delta^2\right)}{648 (m_N-m_\Delta)^3 m_\Delta^3}\\&-\frac{h_A b_{5} (m_N-3 m_\Delta)}{72 (m_N-m_\Delta)^2 m_\Delta}+\frac{h_A b_{4} (-2 m_N+m_\Delta)}{36 (m_N-m_\Delta)^2 m_\Delta}\,, \\
e_{11,\Delta}&=\frac{g_A h_A^2 \left(m_N^2+m_N m_\Delta-6 m_\Delta^2\right)}{72 m_N (m_N-m_\Delta)^2 m_\Delta (m_N+m_\Delta)}\\&-\frac{5 g_1 h_A^2 \left(4 m_N^5+4 m_N^4 m_\Delta+7 m_N^3 m_\Delta^2-5 m_N^2 m_\Delta^3+18 m_N m_\Delta^4-12 m_\Delta^5\right)}{648 m_N (m_N-m_\Delta)^2 m_\Delta^4 (m_N+m_\Delta)}\\&+\frac{b_{4} \left(h_A m_N^2+h_A m_N m_\Delta-3 h_A m_\Delta^2\right)}{36 m_N^3 m_\Delta-36 m_N m_\Delta^3}-\frac{b_{5} \left(h_A m_N^2+h_A m_N m_\Delta-3 h_A m_\Delta^2\right)}{36 m_N^3 m_\Delta-36 m_N m_\Delta^3}\,, \\
e_{12,\Delta}&=\frac{g_A h_A^2 \left(2 m_N^3-7 m_N^2 m_\Delta+13 m_N m_\Delta^2-6 m_\Delta^3\right)}{72 m_N (m_N-m_\Delta)^3 m_\Delta^2}\\&+\frac{5 g_1 h_A^2 \left(4 m_N^5+4 m_N^4 m_\Delta-33 m_N^3 m_\Delta^2+45 m_N^2 m_\Delta^3-42 m_N m_\Delta^4+12 m_\Delta^5\right)}{648 m_N (m_N-m_\Delta)^3 m_\Delta^4}\\&+\frac{h_A b_{4} \left(2 m_N^3-5 m_N^2 m_\Delta+7 m_N m_\Delta^2-3 m_\Delta^3\right)}{36 m_N (m_N-m_\Delta)^2 m_\Delta^2}\\&-\frac{h_A b_{5} \left(2 m_N^3-5 m_N^2 m_\Delta+7 m_N m_\Delta^2-3 m_\Delta^3\right)}{36 m_N (m_N-m_\Delta)^2 m_\Delta^2}\,, \\
e_{13,\Delta}&=\frac{g_A h_A^2 m_N \left(-3 m_N^2+m_\Delta^2\right)}{36 (m_N-m_\Delta)^3 m_\Delta^2 (m_N+m_\Delta)}+\frac{5 g_1 h_A^2 m_N^2 \left(4 m_N^2-6 m_N m_\Delta-3 m_\Delta^2\right)}{162 m_\Delta^3 (-m_N+m_\Delta)^3 (m_N+m_\Delta)}\\&-\frac{h_A b_{4} m_N^3}{18 (m_N-m_\Delta)^2 m_\Delta^2 (m_N+m_\Delta)}+\frac{h_A b_{5} m_N^3}{18 (m_N-m_\Delta)^2 m_\Delta^2 (m_N+m_\Delta)}\,, \\
e_{14,\Delta}&=\frac{h_A^2 \left(2 m_N^2-m_N m_\Delta+3 m_\Delta^2\right)}{72 (m_N-m_\Delta)^2 m_\Delta^2 (m_N+m_\Delta)}\,, \\
e_{15,\Delta}&=\frac{h_A^2 m_N \left(m_N^2-m_N m_\Delta+m_\Delta^2\right)}{9 m_\Delta^2 (-m_N+m_\Delta)^3 (m_N+m_\Delta)}\,, \\
e_{16,\Delta}&=\frac{h_A^2 m_N^3}{9 (m_N-m_\Delta)^3 m_\Delta^2 (m_N+m_\Delta)}\,, \\
e_{17,\Delta}&=-\frac{h_A^2 \left(m_N^2-2 m_N m_\Delta+3 m_\Delta^2\right)}{72 m_N (m_N-m_\Delta)^2 m_\Delta (m_N+m_\Delta)}\,, \\
e_{18,\Delta}&=\frac{h_A^2 m_N \left(m_N^2-4 m_N m_\Delta+m_\Delta^2\right)}{36 m_\Delta^2 (-m_N+m_\Delta)^3 (m_N+m_\Delta)}\,, \\
e_{34,\Delta}&=-\frac{g_A h_A^2 \left(2 m_N^2+6 m_N m_\Delta+5 m_\Delta^2\right)}{72 m_N (m_N-m_\Delta) m_\Delta^2 (m_N+m_\Delta)}\\&-\frac{5 g_1 h_A^2 \left(4 m_N^5+8 m_N^4 m_\Delta-11 m_N^3 m_\Delta^2-19 m_N^2 m_\Delta^3-6 m_N m_\Delta^4+12 m_\Delta^5\right)}{648 m_N (m_N-m_\Delta)^2 m_\Delta^4 (m_N+m_\Delta)}\\&+\frac{h_A b_{4} \left(-2 m_N^3-3 m_N^2 m_\Delta+m_N m_\Delta^2+3 m_\Delta^3\right)}{36 m_\Delta^2 \left(m_N^3-m_N m_\Delta^2\right)}\\&+\frac{h_A b_{5} \left(2 m_N^3+3 m_N^2 m_\Delta-m_N m_\Delta^2-3 m_\Delta^3\right)}{36 m_\Delta^2 \left(m_N^3-m_N m_\Delta^2\right)}\,, \\
\end{aligned}
\end{equation}
and
\begin{equation}
  \label{eq:11a}
  \begin{aligned}
2 e_{19,\Delta}- e_{22,\Delta}-\hat e_{36,\Delta}&=0\,, \\
 e_{20,\Delta}+e_{35,\Delta}&=-\frac{h_A^2 m_N (m_N+2 m_\Delta)}{18 (m_N-m_\Delta)^2 m_\Delta^2 (m_N+m_\Delta)}\,, \\
2e_{21,\Delta}-e_{37,\Delta}&=-\frac{h_A^2}{36 m_N^2 m_\Delta-36 m_N m_\Delta^2}\,, \\
e_{22,\Delta}-4 e_{38,\Delta}&=\frac{h_A^2 (2 m_N+3 m_\Delta)}{36
  m_\Delta^2 \left(m_N^2-m_\Delta^2\right)}\,.\\
 \end{aligned}
\end{equation}
Whereas the contributions of the explicit Roper resonance have the form
\begin{equation}
  \label{eq:11b}
  \begin{aligned}
    c_{1,R}&=0\,, \\
    c_{2,R}&=\frac{g_{RN}^2 m_N}{2 m_N^2-2 m_R^2}\,, \\
    c_{3,R}&=-\frac{g_{RN}^2}{4 (m_N-m_R)}\,, \\
    c_{4,R}&=\frac{g_{RN}^2}{2 m_N-2 m_R}\,, \\
    d_{1+2,R}&=\frac{g_{RN}^2 (3 m_N-m_R)}{8 (m_N-m_R)^2 (m_N+m_R)}\,, \\
d_{3,R}&=-\frac{g_{RN}^2 m_N^2}{2 (m_N-m_R)^2 (m_N+m_R)^2}\,, \\
d_{4,R}&=-\frac{g_A g_{RN}^2}{16 (m_N-m_R)^2}+\frac{g_{R\Delta} g_{RN} h_A}{18 (m_N-m_R) (m_N-m_\Delta)}\\&+\frac{g_{RN}^2 g_{RR}}{16 (m_N-m_R)^2}-\frac{g_{RN} c^R_{4}}{4 (m_N-m_R)}\,, \\
d_{5,R}&=0\,, \\
d_{10,R}&=-\frac{3 g_A g_{RN}^2}{8 (m_N-m_R)^2}+\frac{2 g_{R\Delta} g_{RN} h_A}{3 (m_N-m_R) (m_N-m_\Delta)}\\&+\frac{3 g_{RN}^2 g_{RR}}{8 (m_N-m_R)^2}+\frac{g_{RN} c^R_{3}}{m_N-m_R}-\frac{g_{RN} c^R_{4}}{m_N-m_R}\,, \\
d_{11,R}&=\frac{g_A g_{RN}^2}{4 (m_N-m_R)^2}-\frac{2 g_{R\Delta} g_{RN} h_A}{9 (m_N-m_R) (m_N-m_\Delta)}\\&-\frac{g_{RN}^2 g_{RR}}{4 (m_N-m_R)^2}+\frac{g_{RN} c^R_{4}}{m_N-m_R}\,, \\
d_{12,R}&=\frac{g_A g_{RN}^2 \left(5 m_N^2+2 m_N m_R-m_R^2\right)}{4 \left(m_N^2-m_R^2\right)^2}\\&-\frac{2 g_{R\Delta} g_{RN} h_A m_N \left(2 m_N^2+m_N (2 m_R-m_\Delta)+m_\Delta (m_R+2 m_\Delta)\right)}{9 \left(m_N^2-m_R^2\right) (m_N-m_\Delta) m_\Delta^2}\\&-\frac{g_{RN}^2 g_{RR} m_N (m_N+2 m_R)}{2 \left(m_N^2-m_R^2\right)^2}+\frac{g_{RN} c^R_{2}}{m_N-m_R}+\frac{2 g_{RN} c^R_{4} m_N}{m_N^2-m_R^2}\,, \\
\end{aligned}
\end{equation}
\begin{equation}
\begin{aligned}
d_{13,R}&=-\frac{g_A g_{RN}^2 (3 m_N-m_R)}{4 (m_N-m_R)^2 (m_N+m_R)}-\frac{2 g_{R\Delta} g_{RN} h_A m_N (m_N-m_R-2 m_\Delta)}{9 \left(m_N^2-m_R^2\right) (m_N-m_\Delta) m_\Delta}\\&+\frac{g_{RN}^2 g_{RR} m_N m_R}{\left(m_N^2-m_R^2\right)^2}-\frac{2 g_{RN} c^R_{4} m_N}{m_N^2-m_R^2}\,, \\ 
d_{14-15,R}&=-\frac{g_{RN}^2}{4 (m_N-m_R)^2}\,, \\
d_{16,R}&=\frac{2 g_{RN} c^R_{1}}{m_N-m_R}\,, \\
e_{10,R}&=-\frac{g_{RN}^2 g_{RR}}{8 (m_N-m_R)^3}\\&+\frac{g_{R\Delta} g_{RN} h_A \left(2 m_N^2+m_\Delta (m_R+2 m_\Delta)-m_N (2 m_R+3 m_\Delta)\right)}{36 (m_N-m_R)^2 (m_N-m_\Delta)^2 m_\Delta}\\&+\frac{g_A g_{RN}^2 (5 m_N-m_R)}{32 m_N (m_N-m_R)^3}+\frac{g_{RN} c^R_{4}}{4 (m_N-m_R)^2}\,, \\
e_{11,R}&=-\frac{g_A g_{RN}^2 m_R}{8 m_N (m_N-m_R)^2 (m_N+m_R)}\\&+\frac{g_{R\Delta} g_{RN} h_A \left(m_N^3-m_N^2 (m_R-5 m_\Delta)+3 m_R m_\Delta^2+m_N m_\Delta (-m_R+m_\Delta)\right)}{36 m_N \left(m_N^2-m_R^2\right) m_\Delta \left(m_N^2-m_\Delta^2\right)}\\&+\frac{g_{RN}^2 g_{RR} \left(3 m_N^2+2 m_N m_R+3 m_R^2\right)}{32 m_N \left(m_N^2-m_R^2\right)^2}-\frac{g_{RN} c^R_{4} (m_N+3 m_R)}{8 \left(m_N^3-m_N m_R^2\right)}\,, \\
e_{12,R}&=-\frac{g_{RN} c^R_{4} \left(3 m_N^2-2 m_N m_R+3 m_R^2\right)}{8 m_N (m_N-m_R)^2 (m_N+m_R)}\\&-\frac{g_{R\Delta} g_{RN} h_A }{36 m_N (m_N-m_R)^2 (m_N+m_R) (m_N-m_\Delta)^2 m_\Delta^2} \Big(2 m_N^5-m_N^4 m_\Delta\\&\quad+3 m_R^2 m_\Delta^3-m_N m_R m_\Delta^2 (7 m_R+2 m_\Delta)-m_N^3 \left(2 m_R^2+4 m_R m_\Delta+3 m_\Delta^2\right)\\&\quad+m_N^2 m_\Delta \left(5 m_R^2+6 m_R m_\Delta+3 m_\Delta^2\right)\Big)\\&-\frac{g_A g_{RN}^2 \left(3 m_N^3+m_R^3\right)}{8 m_N (m_N-m_R)^3 (m_N+m_R)^2}+\frac{g_{RN}^2 g_{RR} \left(5 m_N^3+9 m_N^2 m_R-m_N m_R^2+3 m_R^3\right)}{32 m_N (m_N-m_R)^3 (m_N+m_R)^2}\,, \\
e_{13,R}&=\frac{g_A g_{RN}^2 m_N \left(7 m_N^2+2 m_N m_R-m_R^2\right)}{8 \left(m_N^2-m_R^2\right)^3}\\&+\frac{g_{R\Delta} g_{RN} h_A m_N^2 }{18 \left(m_N^2-m_R^2\right)^2 (m_N-m_\Delta)^2 m_\Delta^2 (m_N+m_\Delta)}\Big(m_N^4+m_N^3 (m_R+3 m_\Delta)\\&\quad+m_\Delta^2 \left(m_R^2+2 m_R m_\Delta+4 m_\Delta^2\right)\\&\quad-m_N^2 \left(m_R^2+2 m_R m_\Delta+5 m_\Delta^2\right)-m_N \left(m_R^3+m_R^2 m_\Delta+2 m_\Delta^3\right)\Big)\\&-\frac{g_{RN}^2 g_{RR} m_N^2 (m_N+3 m_R)}{4 \left(m_N^2-m_R^2\right)^3}+\frac{g_{RN} c^R_{4} m_N^2}{\left(m_N^2-m_R^2\right)^2}\,, \\
\end{aligned}
\end{equation}
\begin{equation}
\begin{aligned}
e_{14,R}&=\frac{g_{RN}^2 (3 m_N-m_R)}{64 m_N (m_N-m_R)^2 (m_N+m_R)}\,, \\
e_{15,R}&=-\frac{g_{RN}^2 m_N (2 m_N-m_R)}{8 (m_N-m_R)^3 (m_N+m_R)^2}\,, \\
e_{16,R}&=\frac{g_{RN}^2 m_N^3}{4 (m_N-m_R)^3 (m_N+m_R)^3}\,, \\
e_{17,R}&=-\frac{g_{RN}^2}{32 m_N (m_N-m_R)^2}\,, \\
e_{18,R}&=\frac{g_{RN}^2 m_N}{8 (m_N-m_R)^3 (m_N+m_R)}\,, \\
e_{34,R}&=\frac{3 g_A g_{RN}^2}{32 m_N (m_N-m_R)^2}\\&+\frac{g_{R\Delta} g_{RN} h_A }{36 m_N (m_N-m_R) (m_N+m_R) m_\Delta^2 \left(m_N^2-m_\Delta^2\right)}\Big(2 m_N^4+3 m_N^2 (m_R-m_\Delta) m_\Delta\\&\quad-3 m_R m_\Delta^3-m_N m_\Delta^2 (m_R+3 m_\Delta)+m_N^3 (2 m_R+3 m_\Delta)\Big)\\&-\frac{3 g_{RN}^2 g_{RR}}{32 m_N (m_N-m_R)^2}+\frac{3 g_{RN} c^R_{4}}{8 m_N^2-8 m_N m_R}\,, \\
\end{aligned}
\end{equation}
and
\begin{equation}
\begin{aligned}
e_{20,R}+\hat e_{35,R}&=0\,, \\
2 e_{21,R}- e_{37,R}&=0\,, \\
2 e_{19,R}-e_{22,R}- e_{36,R}&=-\frac{g_{RN}^2}{16 \left(m_N^3-m_N m_R^2\right)}\,, \\
e_{22,R}-e_{34,R}&=\frac{g_{RN}^2}{32 m_N^3-32 m_N m_R^2}\,.\\
 \end{aligned}
\end{equation}

\pagebreak


\section*{}
\begin{table}[h!]
  \centering
  \begin{tabular*}{0.8\textwidth}{@{\extracolsep{\fill}} c| rr| rr|
    rr}\multicolumn{1}{c}{}&\multicolumn{2}{c}{HB-NN} &\multicolumn{2}{c}{HB-$\pi$N}
    &\multicolumn{2}{c}{Cov}\\\hline\hline
    $Q^2$ &\multicolumn{1}{c}{$\pi$N}&\multicolumn{1}{c|}{$\pi$N+RS}
&\multicolumn{1}{c}{$\pi$N}&\multicolumn{1}{c|}{$\pi$N+RS}
&\multicolumn{1}{c}{$\pi$N}&\multicolumn{1}{c}{$\pi$N+RS} \\\hline \hline
$c_1$&-1.82(5)&-1.69(4)&-1.92(5)&-1.60(5)&-2.16(5)&-2.12(5)\\ $c_2$&2.97(9)&3.17(8)&3.12(9)&3.63(9)&2.55(7)&2.65(7)\\ $c_3$&-6.08(6)&-6.07(5)&-6.23(6)&-6.24(5)&-6.23(5)&-6.28(5)\\ $c_4$&4.19(5)&4.61(2)&4.65(4)&5.22(3)&4.32(2)&4.32(2)\\ 
\hline\hline
$\chi^2_{\pi N}$/dof &0.72 &0.69 &0.69 &0.60 &0.67&0.69\\\hline\hline
$\bar\chi^2_{\pi N}$/dof &116 &128 &98 &121 &413&402\\\hline\hline
  \end{tabular*}
  \begin{tabular*}{0.8\textwidth}{@{\extracolsep{\fill}} c| rr| rr| rr}
    $Q^3$ &\multicolumn{1}{c}{$\pi$N}&\multicolumn{1}{c|}{$\pi$N+RS}
&\multicolumn{1}{c}{$\pi$N}&\multicolumn{1}{c|}{$\pi$N+RS}
&\multicolumn{1}{c}{$\pi$N}&\multicolumn{1}{c}{$\pi$N+RS}\\\hline\hline
$c_1$&-1.66(3)&-1.24(2)&-1.62(2)&-1.64(2)&-1.66(2)&-1.55(2)\\ $c_2$&4.10(5)&4.89(5)&3.42(4)&3.51(3)&3.42(3)&3.60(4)\\ $c_3$&-7.11(2)&-7.25(2)&-6.52(2)&-6.63(2)&-6.51(2)&-6.54(2)\\ $c_4$&4.14(5)&4.74(4)&3.89(4)&4.01(4)&3.78(4)&3.86(3)\\ $d_{1+2}$&2.78(5)&3.39(4)&3.89(5)&4.37(4)&4.07(4)&4.09(4)\\ $d_3$&-1.90(8)&-3.47(7)&-2.53(9)&-3.34(7)&-2.43(4)&-2.50(4)\\ $d_5$&-0.64(5)&0.00(4)&-0.79(5)&-0.56(4)&-0.89(4)&-0.86(4)\\ $d_{14-15}$&-7.41(12)&-7.39(13)&-6.94(14)&-7.49(13)&-6.18(10)&-6.05(10)\\ 
\hline\hline
$\chi^2_{\pi N}$/dof &1.04 &1.04 &1.03 &0.83 &0.97 &1.05\\\hline\hline
$\bar\chi^2_{\pi N}$/dof &14.6 &14.1 &13.0 &14.4 &13.5&13.0\\\hline\hline
  \end{tabular*}
  \begin{tabular*}{0.8\textwidth}{@{\extracolsep{\fill}} c| rr| rr| rr}
    $Q^4$ &\multicolumn{1}{c}{$\pi$N}&\multicolumn{1}{c|}{$\pi$N+RS}
&\multicolumn{1}{c}{$\pi$N}&\multicolumn{1}{c|}{$\pi$N+RS}
&\multicolumn{1}{c}{$\pi$N}&\multicolumn{1}{c}{$\pi$N+RS}\\\hline \hline
$c_1$&-0.44(4)&-1.31(8)&0.12(5)&-1.15(8)&-0.81(4)&-0.82(7)\\ $c_2$&4.32(9)&1.88(23)&4.99(13)&2.39(22)&3.87(8)&3.56(16)\\ $c_3$&-4.40(7)&-4.43(9)&-3.09(9)&-4.44(9)&-4.91(9)&-4.59(9)\\ $c_4$&4.07(11)&3.24(17)&3.60(12)&3.45(17)&4.06(10)&3.44(13)\\ $d_{1+2}$&6.51(6)&5.95(9)&5.54(6)&5.60(9)&5.63(4)&5.43(5)\\ $d_3$&-6.21(6)&-5.64(6)&-4.40(4)&-3.84(4)&-4.75(6)&-4.58(8)\\ $d_5$&-0.07(3)&-0.11(4)&-0.45(4)&-0.89(4)&-0.42(3)&-0.40(4)\\ $d_{14-15}$&-12.08(8)&-11.61(9)&-9.42(6)&-9.45(8)&-10.18(6)&-9.94(7)\\ $e_{14}$&-0.39(24)&0.86(29)&-3.23(30)&1.28(32)&-0.85(22)&-0.63(24)\\ $e_{15}$&-6.94(51)&-11.36(81)&-7.98(53)&-13.26(79)&-5.60(39)&-7.33(45)\\ $e_{16}$&1.62(30)&10.73(95)&-0.19(24)&8.29(95)&0.39(17)&1.86(37)\\ $e_{17}$&0.73(40)&-0.66(46)&3.53(41)&-0.73(47)&-1.15(30)&-0.90(32)\\ $e_{18}$&-0.17(52)&4.47(87)&-0.05(56)&4.17(90)&1.60(35)&3.17(45)\\ 
\hline\hline
$\chi^2_{\pi N}$/dof &1.90 &1.92 &1.83&2.04 &1.94&2.07 \\\hline\hline
$\bar\chi^2_{\pi N}$/dof &4.5&4.8 &4.1 &5.9 &4.9 &5.1 \\\hline\hline
  \end{tabular*}
  \caption{LECs determined from fits including $\chi^2_{\mathrm{RS}}$
    as additional constraints at orders $Q^2$, $Q^3$, $Q^4$ in
    comparison with the values given in \cite{Siemens:2016hdi}. The
    values of the $\pi N$ LECs at orders $Q^2$, $Q^3$, $Q^4$ are given in
    units of GeV$^{-1}$, GeV$^{-2}$ and GeV$^{-3}$, respectively.}
\label{tab:FitST}
\end{table}
\newpage

\begin{table}[ht]
  \centering
  \begin{tabular*}{0.8\textwidth}{@{\extracolsep{\fill}} c| r r| r r| r r}
    \multicolumn{1}{c}{}&\multicolumn{2}{c}{HB-NN} &\multicolumn{2}{c}{HB-$\pi$N}
    &\multicolumn{2}{c}{Cov}\\\hline\hline
$Q^2$ &\multicolumn{1}{c}{$\pi$N} &\multicolumn{1}{c|}{$\pi$N+$\pi\pi$N}&\multicolumn{1}{c}{$\pi$N} &\multicolumn{1}{c|}{$\pi$N+$\pi\pi$N}&\multicolumn{1}{c}{$\pi$N} &\multicolumn{1}{c}{$\pi$N+$\pi\pi$N} \\\hline\hline
$c_1$&-1.69(4)&-1.69(4)&-1.60(5)&-1.59(5)&-2.19(5)&-2.12(5)\\ $c_2$&3.18(8)&3.17(8)&3.63(9)&3.65(9)&2.52(7)&2.65(7)\\ $c_3$&-6.08(5)&-6.07(5)&-6.24(5)&-6.25(6)&-6.25(6)&-6.28(5)\\ $c_4$&4.61(2)&4.61(2)&5.22(3)&5.27(4)&4.32(2)&4.32(2)\\ 
\hline\hline
$\chi^2_{\pi N}$/dof &0.72 &0.72 &0.69 &0.69 &0.67 &0.67\\\hline\hline
$\bar\chi^2_{\pi N}$/dof &116 &116 &98 &97 &413 &415\\\hline\hline
$\chi^2_{\pi\pi N}$/dof &- &1.03 &-  &0.95 &-  &1.09\\\hline\hline
$\bar\chi^2_{\pi\pi N}$/dof &-&34 &-&27 &- &5.5\\\hline\hline
  \end{tabular*}
  \begin{tabular*}{0.8\textwidth}{@{\extracolsep{\fill}} c| r r| r r| r r}
    $Q^3$ &\multicolumn{1}{c}{$\pi$N} &\multicolumn{1}{c|}{$\pi$N+$\pi\pi$N}&\multicolumn{1}{c}{$\pi$N} &\multicolumn{1}{c|}{$\pi$N+$\pi\pi$N}&\multicolumn{1}{c}{$\pi$N} &\multicolumn{1}{c}{$\pi$N+$\pi\pi$N} \\\hline\hline
$c_1$&-1.24(2)&-1.24(2)&-1.64(2)&-1.64(2)&-1.55(2)&-1.55(2)\\ $c_2$&4.89(5)&4.89(5)&3.51(3)&3.51(3)&3.60(4)&3.60(4)\\ $c_3$&-7.25(2)&-7.26(2)&-6.63(2)&-6.63(2)&-6.54(2)&-6.54(2)\\ $c_4$&4.74(4)&4.74(4)&4.01(4)&4.01(4)&3.86(3)&3.86(3)\\ $d_{1+2}$&3.39(4)&3.39(4)&4.37(4)&4.37(4)&4.09(4)&4.09(4)\\ $d_3$&-3.47(7)&-3.44(7)&-3.34(7)&-3.35(7)&-2.50(4)&-2.50(4)\\ $d_4$&-&3.7(2.3)&-&3.1(2.2)&-&3.3(2.1)\\ $d_5$&0.00(4)&-0.02(4)&-0.56(4)&-0.56(4)&-0.86(4)&-0.85(4)\\ $d_{10}$&-&10.9(5.6)&-&-0.8(4.9)&-&-6.4(4.6)\\ $d_{11}$&-&-30.9(7.6)&-&-15.6(6.7)&-&-1.7(6.6)\\ $d_{12}$&-&-10.9(6.0)&-&5.9(5.4)&-&11.6(4.7)\\ $d_{13}$&-&27.7(7.7)&-&13.6(6.8)&-&-1.9(6.4)\\ $d_{14-15}$&-7.39(13)&-7.36(13)&-7.49(13)&-7.43(13)&-6.05(10)&-6.02(10)\\ $d_{16}$&-&-3.0(1.6)&-&0.4(1.3)&-&0.5(1.1)\\ $l_{1}$&-&-0.39(60)&-&-0.39(60)&-&-0.35(60)\\ $l_{2}$&-&4.30(10)&-&4.29(10)&-&4.30(10)\\ $l_{3}$&-&3.0(2.4)&-&3.2(2.4)&-&3.2(2.4)\\ $l_{4}$&-&4.40(20)&-&4.41(20)&-&4.40(20)\\ 
\hline\hline
$\chi^2_{\pi N}$/dof &1.04 &1.01 &1.03 &1.00 &0.97 &0.97\\\hline\hline
$\bar\chi^2_{\pi N}$/dof &14.6 &14.6 &13.0 &13.1 &13.5 &13.6\\\hline\hline
$\chi^2_{\pi\pi N}$/dof &- &0.72 &- &1.00 &- &0.96\\\hline\hline
$\bar\chi^2_{\pi\pi N}$/dof &- &5.3 &- &6.5 &- &8.0\\\hline\hline
  \end{tabular*}
  \caption{LECs determined from fits at orders $Q^2$ and $Q^3$ with
    additional constraints from the reaction $\pi N\to\pi\pi N$ with
    $T_{\pi,\pi\pi N}<275$~MeV.  The
    values of the $\pi N$ LECs at orders $Q^2$ and $Q^3$ are given in
    units of GeV$^{-1}$ and GeV$^{-2}$, respectively,
    while the $l_i$'s are dimensionless.}
\label{tab:FitpipiN1}
\end{table}
\newpage
\begin{table}[ht]
  \centering
  \begin{tabular*}{0.8\textwidth}{@{\extracolsep{\fill}} c| r r| r r|
    r r}
\multicolumn{1}{c}{}&\multicolumn{2}{c}{HB-NN} &\multicolumn{2}{c}{HB-$\pi$N}
    &\multicolumn{2}{c}{Cov}\\\hline\hline
    $Q^4$ &\multicolumn{1}{c}{$\pi$N} &\multicolumn{1}{c|}{$\pi$N+$\pi\pi$N}&\multicolumn{1}{c}{$\pi$N} &\multicolumn{1}{c|}{$\pi$N+$\pi\pi$N}&\multicolumn{1}{c}{$\pi$N} &\multicolumn{1}{c}{$\pi$N+$\pi\pi$N} \\\hline\hline
$c_1$&-1.31(8)&-1.06(6)&-1.15(8)&-1.03(6)&-0.82(7)&-0.89(6)\\ $c_2$&1.88(23)&2.44(17)&2.39(22)&2.52(18)&3.56(16)&3.38(15)\\ $c_3$&-4.43(9)&-4.29(9)&-4.44(9)&-4.24(9)&-4.59(9)&-4.59(9)\\ $c_4$&3.24(17)&3.10(15)&3.45(17)&3.03(15)&3.44(13)&3.31(13)\\ $d_{1+2}$&5.95(9)&5.85(8)&5.60(9)&5.35(8)&5.43(5)&5.40(5)\\ $d_3$&-5.64(6)&-5.58(6)&-3.84(4)&-3.76(4)&-4.58(8)&-4.60(7)\\ $d_4$&-&-1.8(1.5)&-&-1.5(1.2)&-&-4.3(1.5)\\ $d_5$&-0.11(4)&-0.08(4)&-0.89(4)&-0.80(4)&-0.40(4)&-0.37(4)\\ $d_{10}$&-&-24.7(3.7)&-&-29.6(2.2)&-&-31.9(2.6)\\ $d_{11}$&-&2.9(5.0)&-&13.1(3.0)&-&20.5(3.9)\\ $d_{12}$&-&25.2(3.9)&-&28.1(2.2)&-&34.0(2.7)\\ $d_{13}$&-&-7.1(5.1)&-&-16.3(3.0)&-&-24.2(3.7)\\ $d_{14-15}$&-11.61(9)&-11.51(9)&-9.45(8)&-9.24(7)&-9.94(7)&-9.88(7)\\ $d_{16}$&-&4.11(96)&-&9.16(85)&-&0.82(80)\\ $e_{10}$&-&-34.4(8.1)&-&-33.8(7.9)&-&-22.8(6.3)\\ $e_{11}$&-&4.4(4.5)&-&13.9(5.2)&-&3.6(5.4)\\ $e_{12}$&-&56.1(4.3)&-&53.2(3.9)&-&23.5(3.9)\\ $e_{13}$&-&-57.5(7.1)&-&-61.9(7.7)&-&-19.8(6.4)\\ $e_{14}$&0.86(29)&0.81(29)&1.28(32)&1.35(31)&-0.63(24)&-0.58(24)\\ $e_{15}$&-11.36(81)&-11.39(78)&-13.26(79)&-14.11(77)&-7.33(45)&-7.48(45)\\ $e_{16}$&10.73(95)&9.15(78)&8.29(95)&8.38(81)&1.86(37)&2.22(36)\\ $e_{17}$&-0.66(46)&-0.80(46)&-0.73(47)&-1.01(47)&-0.90(32)&-0.83(32)\\ $e_{18}$&4.47(87)&5.20(81)&4.17(90)&6.14(82)&3.17(45)&3.49(44)\\ $e_{34}$&-&-0.9(15.5)&-&-11.8(18.0)&-&3.8(21.6)\\ $l_{1}$&-&-0.36(60)&-&-0.40(60)&-&-0.29(60)\\ $l_{2}$&-&4.29(10)&-&4.29(10)&-&4.29(10)\\ $l_{3}$&-&3.1(2.4)&-&2.9(2.4)&-&3.3(2.4)\\ $l_{4}$&-&4.42(20)&-&4.42(20)&-&4.39(20)\\ 
\hline\hline
$\chi^2_{\pi N}$/dof &1.90 &1.90 &1.83 &1.83 &1.94 &1.90\\\hline\hline
$\bar\chi^2_{\pi N}$/dof &4.5 &4.6 &4.1 &4.1 &4.9 &4.9\\\hline\hline
$\chi^2_{\pi\pi N}$/dof &- &2.1 &- &2.8 &- &2.5\\\hline\hline
$\bar\chi^2_{\pi\pi N}$/dof &- &12 &- &17 &- &6.3\\\hline\hline
  \end{tabular*}
\caption{LECs determined from fits at order $Q^4$ with
    additional constraints from the reaction $\pi N\to\pi\pi N$ with
    $T_{\pi,\pi\pi N}<275$~MeV. The
    values of the $\pi N$ LECs at orders $Q^2$, $Q^3$, $Q^4$ are given in
    units of GeV$^{-1}$, GeV$^{-2}$ and GeV$^{-3}$, respectively,
    while the $l_i$'s are dimensionless.}
\label{tab:FitpipiN2}
\end{table}

\newpage
\newcolumntype{C}{>{\centering\arraybackslash}p{2.5em}}
\begin{table}[h]
  \centering
  \begin{tabular*}{0.4\textwidth}{@{\extracolsep{\fill}} C| r||  C| r}\hline\hline
$d_4$&-5.4(1.7) &$e_{10}$ &-9.3(8.4)\\
$d_{10}$&-33.2(5.8) &$e_{11}$ &4.9(6.3)\\
$d_{11}$&18.2(8.5) &$e_{12}$ &7.3(4.6)\\
$d_{12}$ &30.8(6.2) &$e_{13}$ &-11.4(6.6)\\
$d_{13}$&-20.0(8.3) &$e_{34}$ &-6.7(20.9)\\
$d_{16}$&1.7(1.0) &\\
\hline\hline
  \end{tabular*}\hfill 
  \begin{tabular*}{0.4\textwidth}{@{\extracolsep{\fill}} c|| c c c c} \hline\hline
&$d_{10}$&$d_{11}$&$d_{12}$&$d_{13}$\\ \hline\hline
$d_{10}$ & &-70 &-97 &67\\ 
$d_{11}$ &-89 & &72 &-99\\ 
$d_{12}$ &-99 &88 & &-73\\ 
$d_{13}$ &91 &-100 &-90 &\\
\hline\hline
  \end{tabular*}
  \caption{The left table shows 3$\pi$NN LECs determined from
    covariant fits at order $Q^4+\delta^1$ with $T_{\pi,\pi\pi N}<275$~MeV.
The upper/lower triangle in the right table correspond to a selected
part of
    the correlation matrix for the covariant fits at
    $Q^4$/$Q^4+\delta^1$. The
    values of the LECs $d_i$ and $e_i$ are given in
    units of GeV$^{-2}$ and GeV$^{-3}$, respectively.}
\label{tab:Q4pipiNLECsCorr}
\end{table}

\begin{table}[h!]
  \centering
  \begin{tabular*}{1\textwidth}{@{\extracolsep{\fill}} c| r r r| c| c}\hline\hline
    $Q^4$ &\multicolumn{1}{c}{Cov} &\multicolumn{1}{c}{$\Delta x$}
    &\multicolumn{1}{c|}{$\overline{\Delta x}$}
    &\multicolumn{1}{c|}{$x_\Delta(h_A\pm g_1\pm b_4\pm b_5)$}&\multicolumn{1}{c}{$x_R(g_{RN}\pm g_{R\Delta}\pm g_{RR}\pm c^R_{1,2,3,4})$}\\\hline\hline
$c_{1}$&-0.89(6)&0.03 &-32&0.01&0\\ 
$c_{2}$&3.38(15)&-1.10&-3.1&-1.78&-0.05\\ 
$c_{3}$&-4.59(9)&0.89&-5.2&2.76&0.06\\ 
$c_{4}$&3.31(13)&-2.59&-1.3&-1.37&-0.12\\ 
$d_{1+2}$&5.40(5)&1.75&3.1&-2.20&0.04\\ 
$d_3$&-4.60(7)&-1.36&3.4&1.36&-0.04\\ 
$d_4$&-4.3(1.5)&-&-&-0.76 $\pm$ 3.71  $\pm$ 0.26  $\mp$ 0.26 &-0.04 $\pm$ 0.18  $\pm$ 0.03  $\pm$ 0.17 \\ 
$d_5$&-0.37(4)&-0.55&0.7&0.35&0\\ 
$d_{10}$&-31.9(2.6)&-18.6&1.7&-9.08 $\pm$ 5.50  $\pm$ 3.06  $\pm$ 2.55 &-0.24 $\pm$ 2.14  $\pm$ 0.18  $\mp$ 0.70  $\pm$ 0.70 \\ 
$d_{11}$&20.5(3.9)&7.01&2.9&3.03 $\mp$ 4.34  $\mp$ 1.02  $\mp$ 0.51 &0.16 $\mp$ 0.71  $\mp$ 0.12  $\mp$ 0.70 \\ 
$d_{12}$&34.0(2.7)&16.8&2.0&5.81 $\mp$ 4.47  $\mp$ 1.96  $\mp$ 1.76 &0.14 $\mp$ 1.50  $\mp$ 0.15  $\mp$ 0.70  $\mp$ 0.55 \\ 
$d_{13}$&-24.2(3.7)&-6.41&3.8&-2.30 $\pm$ 3.52  $\pm$ 0.78  $\pm$ 0.57 &-0.09 $\pm$ 0.68  $\pm$ 0.12  $\pm$ 0.55 \\ 
$d_{14-15}$&-9.88(7)&-1.31&7.6&3.57&-0.12\\ 
$d_{16}$&0.82(80)&-3.95&-0.2&0&-1.40 \\ 
$e_{10}$&-22.8(6.3)&22.2&-1.0&1.65 $\mp$ 10.99  $\mp$ 0.23  $\pm$ 0.49 &-0.14 $\pm$ 0.20  $\pm$ 0.12  $\pm$ 0.35 \\ 
$e_{11}$&3.6(5.4)&-1.41&-2.6&-1.07 $\mp$ 0.97  $\pm$ 0.13  $\mp$ 0.13 &-0.05 $\pm$ 0.17  $\pm$ 0.03  $\pm$ 0.21 \\ 
$e_{12}$&23.5(3.9)&-30.7&-0.8&-1.22 $\pm$ 10.21  $\pm$ 0.18  $\mp$ 0.18 &0.16 $\mp$ 0.19  $\mp$ 0.13  $\mp$ 0.48 \\ 
$e_{13}$&-19.8(6.4)&35.4&-0.6&0.82 $\mp$ 8.82  $\mp$ 0.22  $\pm$ 0.22 &-0.07 $\pm$ 0.14  $\pm$ 0.08  $\pm$ 0.22 \\ 
$e_{14}$&-0.58(24)&-1.88&0.3&0.46&0\\ 
$e_{15}$&-7.48(45)&-2.69&2.8&2.83&0\\ 
$e_{16}$&2.22(36)&4.48&0.5&-2.00&-0.01\\ 
$e_{17}$&-0.83(32)&-0.29&2.8&-0.37&-0.02\\ 
$e_{18}$&3.49(44)&7.23&0.5&-1.27&-0.05\\ 
$e_{34}$&3.8(21.6)&-1.83&-2.1&0.59 $\pm$ 1.07  $\mp$ 0.09  $\pm$ 0.09 &0.06 $\mp$ 0.09  $\mp$ 0.05  $\mp$ 0.28 \\ 
\hline\hline
  \end{tabular*}
\caption{LECs determined from fits at order $Q^4$ in the covariant
  approach with the
    additional constraints from the reaction $\pi N\to\pi\pi N$ along
    with the RG-quantities $\Delta x$ and $\overline{\Delta x}$
    defined in Eq.~\eqref{eq:9}.  $x_\Delta$ and $x_R$ denote
    the saturations of the LECs by the $\Delta$ and Roper resonances,
    respectively, using $h_A=1.35$, $g_1=\pm2.29$, $g_{RN}=0.35$,
  $b_i=g_{R\Delta}=g_{RR}=c_i^R=\pm1$. The
    values of the $\pi N$ LECs at orders $Q^2$, $Q^3$, $Q^4$ are given in
    units of GeV$^{-1}$, GeV$^{-2}$ and GeV$^{-3}$, respectively.}
\label{tab:Q4SatFlow}
\end{table}
\newpage

\newcolumntype{C}{>{\centering\arraybackslash}p{6.5em}}
\begin{table}[h]
  \centering
  \begin{tabular*}{0.8\textwidth}{@{\extracolsep{\fill}} c| C |  C|
    C| c}\hline\hline
   $Q^4$  &\multicolumn{1}{c}{$\pi$N} &\multicolumn{1}{c}{$\pi$N+RS} &\multicolumn{1}{c}{$\pi$N+$\pi\pi$N} &\multicolumn{1}{c}{RS} \\\hline\hline
$d_{00}^+[M_\pi^{-1}]$&-0.37(12)(46)&-1.60(3)(3)&-0.61(10)(40)&-1.36(3)\\ $d_{10}^+[M_\pi^{-3}]$&-0.86(20)(71)&1.14(6)(13)&-0.51(17)(63)&1.16(2)\\ $d_{01}^+[M_\pi^{-3}]$&0.79(4)(22)&0.92(3)(18)&0.76(4)(23)&1.16(2)\\ $d_{20}^+[M_\pi^{-5}]$&1.29(9)(25)&0.39(3)(4)&1.14(8)(21)&0.196(3)\\ $d_{11}^+[M_\pi^{-5}]$&0.64(4)(13)&0.42(3)(8)&0.64(4)(13)&0.185(3)\\ $d_{02}^+[M_\pi^{-5}]$&0.033(7)(2)&0.001(6)(8)&0.032(7)(2)&0.0336(6)\\ $b_{00}^+[M_\pi^{-3}]$&-5.2(2)(1.1)&-4.7(1)(1.3)&-5.4(2)(1.1)&-3.45(7)\\ $d_{00}^-[M_\pi^{-2}]$&1.15(2)(15)&1.39(1)(2)&1.23(2)(13)&1.41(1)\\ $d_{10}^-[M_\pi^{-4}]$&0.30(3)(23)&-0.10(2)(7)&0.16(3)(20)&-0.159(4)\\ $d_{01}^-[M_\pi^{-4}]$&-0.210(4)(33)&-0.22(0)(2)&-0.21(0)(3)&-0.141(5)\\ $b_{00}^-[M_\pi^{-2}]$&6.4(7)(2.1)&10.4(4)(5)&5.8(7)(2.2)&10.49(11)\\ $b_{10}^-[M_\pi^{-4}]$&5.8(5)(1.1)&3.1(3)(5)&6.2(5)(1.2)&1.00(3)\\ $b_{01}^-[M_\pi^{-4}]$&0.38(16)(4)&-0.09(14)(7)&0.43(16)(5)&0.21(2)\\ 
\hline\hline
  \end{tabular*}
\vskip 5pt
  \begin{tabular*}{0.8\textwidth}{@{\extracolsep{\fill}} c| C |  C|
    C| c}\hline\hline
$d_{00}^+[M_\pi^{-1}]$&-0.48(12)(22)&-1.69(3)(7)&-0.50(10)(22)&-1.36(3)\\ $d_{10}^+[M_\pi^{-3}]$&-0.67(20)(46)&1.17(5)(4)&-0.68(17)(46)&1.16(2)\\ $d_{01}^+[M_\pi^{-3}]$&0.70(4)(20)&0.73(3)(18)&0.63(4)(21)&1.16(2)\\ $d_{20}^+[M_\pi^{-5}]$&1.30(9)(25)&0.45(2)(5)&1.31(8)(25)&0.196(3)\\ $d_{11}^+[M_\pi^{-5}]$&0.80(4)(17)&0.54(3)(11)&0.85(4)(18)&0.185(3)\\ $d_{02}^+[M_\pi^{-5}]$&0.052(8)(4)&-0.06(1)(2)&0.055(8)(5)&0.0336(6)\\ $b_{00}^+[M_\pi^{-3}]$&-1.44(21)(2.04)&-3.0(2)(1.5)&-2.0(2)(1.9)&-3.45(7)\\ $d_{00}^-[M_\pi^{-2}]$&0.71(2)(24)&1.27(2)(7)&0.79(2)(22)&1.41(1)\\ $d_{10}^-[M_\pi^{-4}]$&0.77(3)(34)&-0.08(3)(10)&0.66(3)(31)&-0.159(4)\\ $d_{01}^-[M_\pi^{-4}]$&-0.060(4)(89)&-0.11(0)(7)&-0.07(0)(9)&-0.141(5)\\ $b_{00}^-[M_\pi^{-2}]$&6.7(8)(1.3)&10.1(5)(6)&4.9(7)(1.7)&10.49(11)\\ $b_{10}^-[M_\pi^{-4}]$&6.3(5)(1.2)&3.6(3)(6)&7.4(5)(1.5)&1.00(3)\\ $b_{01}^-[M_\pi^{-4}]$&0.47(16)(6)&-0.96(14)(27)&0.57(16)(9)&0.21(2)\\ 
\hline\hline
  \end{tabular*}
\vskip 5pt
  \begin{tabular*}{0.8\textwidth}{@{\extracolsep{\fill}} c| C |  C|
    C| c}\hline\hline
$d_{00}^+[M_\pi^{-1}]$&-1.22(9)(12)&-1.46(3)(2)&-1.12(8)(14)&-1.36(3)\\ $d_{10}^+[M_\pi^{-3}]$&0.75(11)(25)&1.14(4)(13)&0.63(11)(28)&1.16(2)\\ $d_{01}^+[M_\pi^{-3}]$&0.97(3)(16)&1.10(3)(13)&0.97(3)(17)&1.16(2)\\ $d_{20}^+[M_\pi^{-5}]$&0.54(4)(11)&0.40(2)(8)&0.58(4)(12)&0.196(3)\\ $d_{11}^+[M_\pi^{-5}]$&0.43(2)(9)&0.34(2)(7)&0.44(2)(10)&0.185(3)\\ $d_{02}^+[M_\pi^{-5}]$&-0.004(6)(5)&-0.012(5)(7)&-0.002(6)(5)&0.0336(6)\\ $b_{00}^+[M_\pi^{-3}]$&-6.05(10)(0.45)&-5.6(1)(6)&-6.10(9)(43)&-3.45(7)\\ $d_{00}^-[M_\pi^{-2}]$&1.40(1)(3)&1.37(1)(3)&1.41(1)(3)&1.41(1)\\ $d_{10}^-[M_\pi^{-4}]$&-0.21(1)(5)&-0.18(1)(5)&-0.21(1)(5)&-0.159(4)\\ $d_{01}^-[M_\pi^{-4}]$&-0.247(3)(23)&-0.24(0)(2)&-0.25(0)(2)&-0.141(5)\\ $b_{00}^-[M_\pi^{-2}]$&8.0(5)(1.3)&10.4(4)(7)&7.6(5)(1.5)&10.49(11)\\ $b_{10}^-[M_\pi^{-4}]$&4.13(27)(88)&3.2(2)(7)&4.31(27)(92)&1.00(3)\\ $b_{01}^-[M_\pi^{-4}]$&0.38(11)(7)&0.44(10)(9)&0.36(11)(7)&0.21(2)\\ 
\hline\hline
  \end{tabular*}
  \caption{Subthreshold parameters at order
    $Q^4$ in comparison with the RS analysis values. The upper/middle/lower table
    refer to results in the HB-NN, HB-$\pi$N and covariant counting,
    respectively. The statistical and
    theoretical uncertainties are given in the first and second
    bracket, respectively.}
\label{tab:SubThrPara}
\end{table}
\newpage

\newcolumntype{C}{>{\centering\arraybackslash}p{7em}}
\begin{table}[h]
  \centering
  \begin{tabular*}{0.8\textwidth}{@{\extracolsep{\fill}} c| C |  C|
    C| c}\hline\hline
   $Q^4$  &\multicolumn{1}{c}{$\pi$N} &\multicolumn{1}{c}{$\pi$N+RS} &\multicolumn{1}{c}{$\pi$N+$\pi\pi$N} &\multicolumn{1}{c}{RS} \\\hline\hline
$a_{0+}^+[M_\pi^{-1}]$&3.1(9)(10)&-6.2(5)(4.5)&0.2(7)(1.1)&-0.9(1.4)\\ $a_{0+}^-[M_\pi^{-1}]$&82.8(3)(5)&83.7(2)(1.6)&82.9(3)(5)&85.4(9)\\ $a_{1+}^+[M_\pi^{-3}]$&136.0(5)(3.9)&137.0(4)(3.2)&135.8(5)(3.9)&131.2(1.7)\\ $a_{1+}^-[M_\pi^{-3}]$&-83.9(5)(2.1)&-87.0(4)(7)&-83.6(5)(2.2)&-80.3(1.1)\\ $a_{1-}^+[M_\pi^{-3}]$&-57.9(7)(3.2)&-56.2(5)(3.1)&-58.7(7)(3.3)&-50.9(1.9)\\ $a_{1-}^+[M_\pi^{-3}]$&-11.9(1.3)(1.9)&-6.2(8)(1.7)&-13.0(1.2)(2.0)&-9.9(1.2)\\ $b_{0+}^+[M_\pi^{-3}]$&-53.5(3.5)(4.1)&-14.4(1.6)(16.4)&-42.7(2.6)(4.7)&-45.0(1.0)\\ $b_{0+}^-[M_\pi^{-3}]$&17.5(4)(1.2)&15.1(4)(4.0)&17.4(5)(1.2)&4.9(8)\\ 
\hline\hline
  \end{tabular*}
\vskip 5pt
  \begin{tabular*}{0.8\textwidth}{@{\extracolsep{\fill}} c| C |  C|
    C| c}\hline\hline
$a_{0+}^+[M_\pi^{-1}]$&2.9(9)(8)&-12.1(6)(4.1)&1.3(7)(1.2)&-0.9(1.4)\\ $a_{0+}^-[M_\pi^{-1}]$&82.2(3)(2)&85.4(3)(1.3)&82.4(3)(3)&85.4(9)\\ $a_{1+}^+[M_\pi^{-3}]$&133.1(5)(4.5)&129.6(4)(4.6)&132.6(5)(4.6)&131.2(1.7)\\ $a_{1+}^-[M_\pi^{-3}]$&-82.7(5)(2.2)&-85.6(4)(9)&-81.6(5)(2.5)&-80.3(1.1)\\ $a_{1-}^+[M_\pi^{-3}]$&-54.1(7)(2.5)&-63.1(6)(2.9)&-55.9(7)(2.6)&-50.9(1.9)\\ $a_{1-}^+[M_\pi^{-3}]$&-10.9(1.3)(2.3)&-10.0(8)(1.8)&-14.1(1.2)(2.6)&-9.9(1.2)\\ $b_{0+}^+[M_\pi^{-3}]$&-52.7(3.5)(2.9)&6.0(2.0)(16.5)&-47.9(2.7)(4.0)&-45.0(1.0)\\ $b_{0+}^-[M_\pi^{-3}]$&22.4(5)(3)&14.9(4)(2.4)&22.1(4)(2)&4.9(8)\\ 
\hline\hline
  \end{tabular*}
\vskip 5pt
  \begin{tabular*}{0.8\textwidth}{@{\extracolsep{\fill}} c| C | C|
    C| c}\hline\hline
$a_{0+}^+[M_\pi^{-1}]$&0.0(9)(1.7)&0.0(5)(2.4)&0.7(8)(1.5)&-0.9(1.4)\\ $a_{0+}^-[M_\pi^{-1}]$&83.3(3)(5)&83.2(2)(5)&83.5(3)(5)&85.4(9)\\ $a_{1+}^+[M_\pi^{-3}]$&135.8(5)(3.5)&137.4(5)(3.0)&135.6(5)(3.5)&131.2(1.7)\\ $a_{1+}^-[M_\pi^{-3}]$&-84.3(5)(1.6)&-86.4(4)(1.0)&-83.9(5)(1.7)&-80.3(1.1)\\ $a_{1-}^+[M_\pi^{-3}]$&-59.6(7)(3.0)&-56.2(6)(2.8)&-60.0(6)(3.1)&-50.9(1.9)\\ $a_{1-}^+[M_\pi^{-3}]$&-13.6(1.2)(2.5)&-7.7(9)(2.2)&-14.8(1.1)(2.6)&-9.9(1.2)\\ $b_{0+}^+[M_\pi^{-3}]$&-37.8(3.5)(6.9)&-35.3(2.0)(9.4)&-41.1(3.2)(6.2)&-45.0(1.0)\\ $b_{0+}^-[M_\pi^{-3}]$&16.3(6)(1.6)&15.8(5)(1.9)&16.0(6)(1.7)&4.9(8)\\ 
\hline\hline
  \end{tabular*}
\caption{Threshold parameters at order
    $Q^4$ in comparison with RS analysis values. The upper/middle/lower table
    refer to results in the HB-NN, HB-$\pi$N and covariant counting,
    respectively.    The statistical and
    theoretical uncertainties are given in the first and second
    bracket, respectively.}
\label{tab:ThrPara}
\end{table}
\clearpage
\section*{}

\vspace{1.5cm}

\clearpage
\newpage

\begin{figure}[ht]
  \centering
\includegraphics[width=0.45\textwidth]{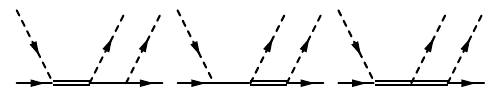}
  \caption{Leading-order $\Delta$ pole diagrams, where the double
    solid line refers to the $\Delta$. For notation see Fig.~\ref{fig:TreeGraphs}.}
  \label{fig:piNRuleDelta}
\end{figure}

\newpage
\begin{figure}[ht]
  \centering
\includegraphics[width=\textwidth]{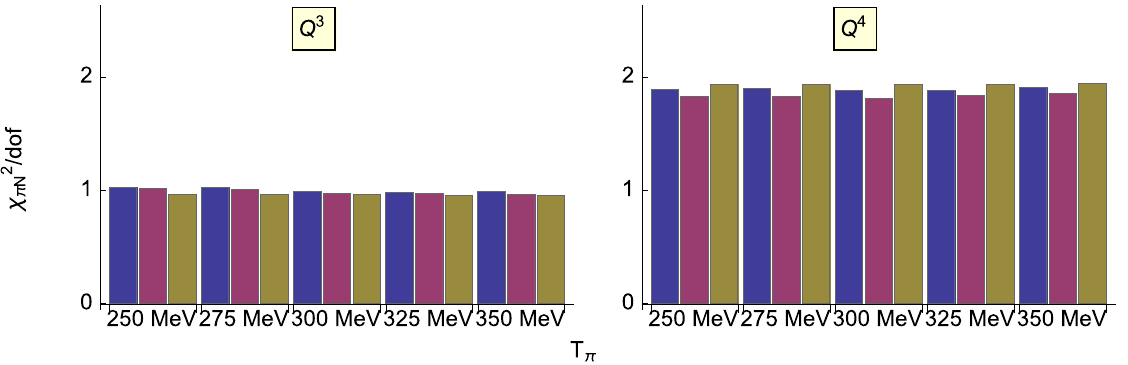}
\includegraphics[width=\textwidth]{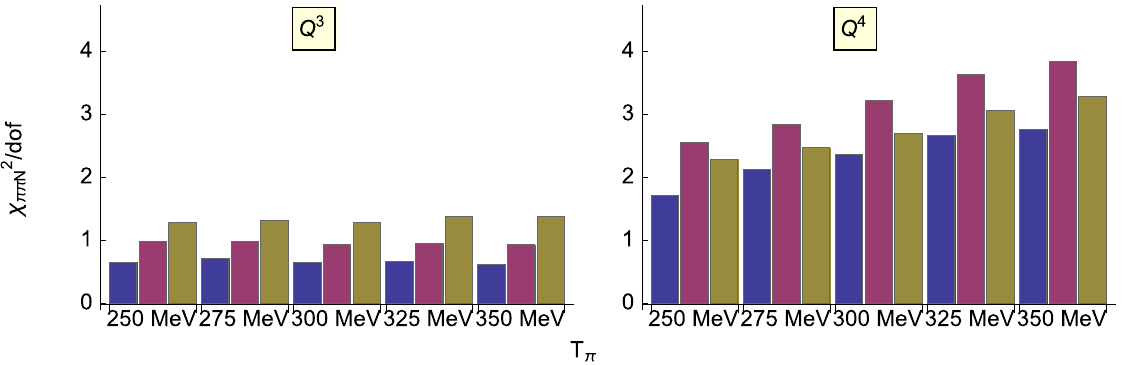}
\includegraphics[width=\textwidth]{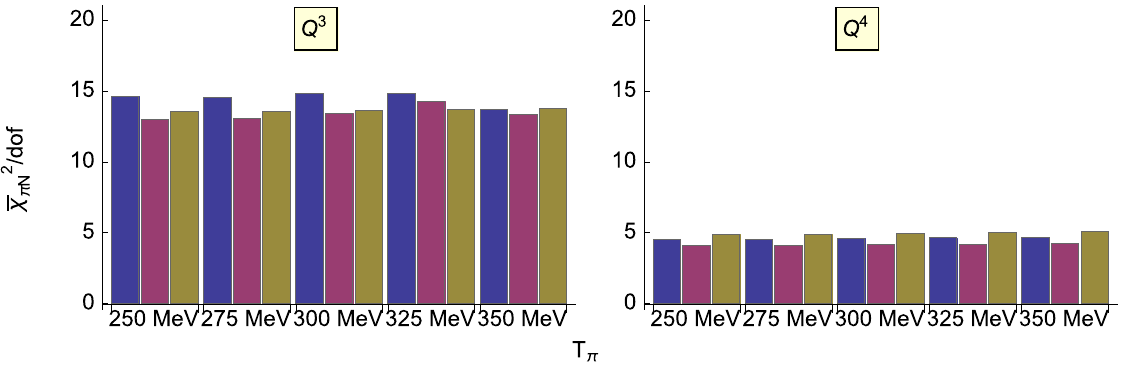}
\includegraphics[width=\textwidth]{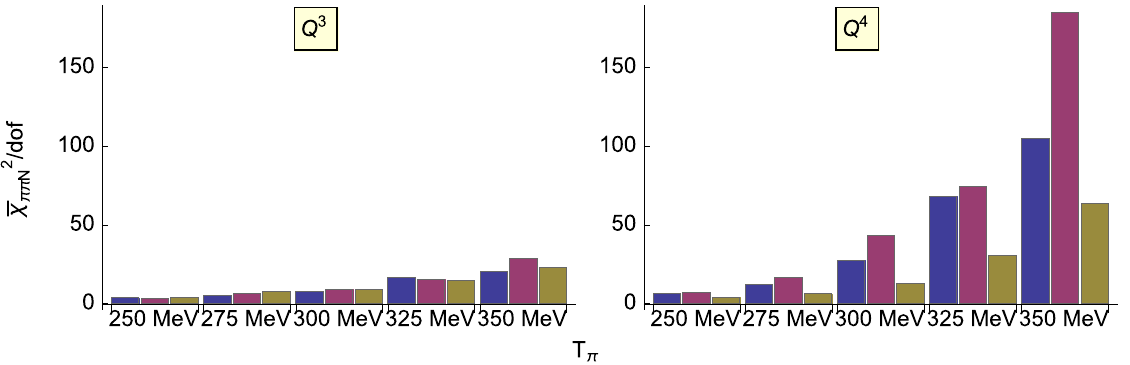}
\caption{Reduced $\chi_{\pi N}^2$/$\chi_{\pi\pi N}^2$ (with theoretical error) and 
        $\bar\chi_{\pi N}^2$/$\bar\chi_{\pi\pi N}^2$
    (without theoretical error) for fits up to various maximum energies
    $T_{\pi,\pi\pi N}$. The blue/red/green bars denote the results for the 
    HB-NN/HB-$\pi$N/Cov counting.}
  \label{fig:RedChiSq}
\end{figure}

\begin{figure}[ht]
  \centering
\includegraphics[width=0.7\textwidth]{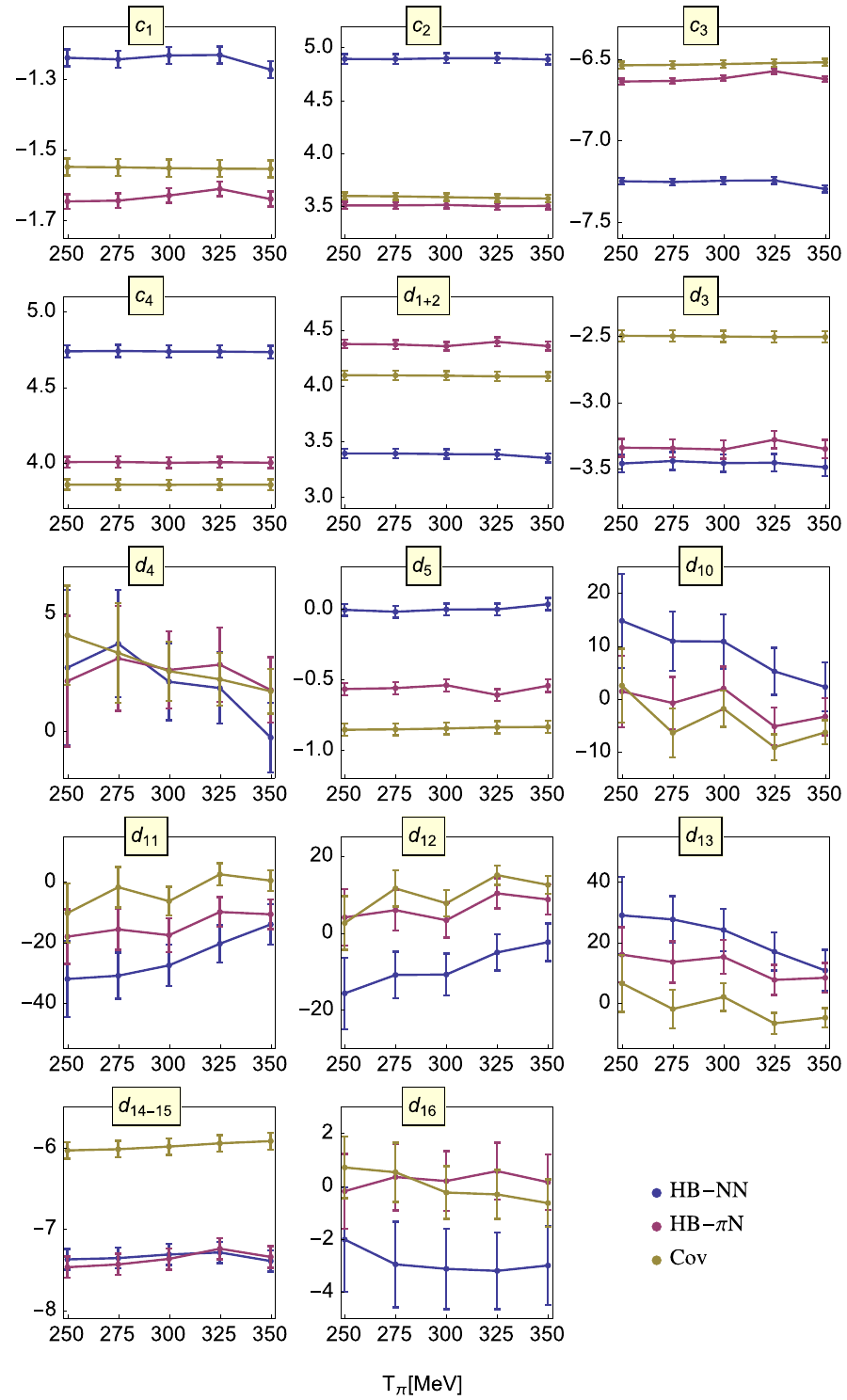}
  \caption{Change of the LECs at $Q^3$ over the maximum fit energy $T_{\pi,\pi\pi N}$.}
  \label{fig:LECsQ3}
\end{figure}

\begin{figure}[ht]
  \centering
\includegraphics[width=0.7\textwidth]{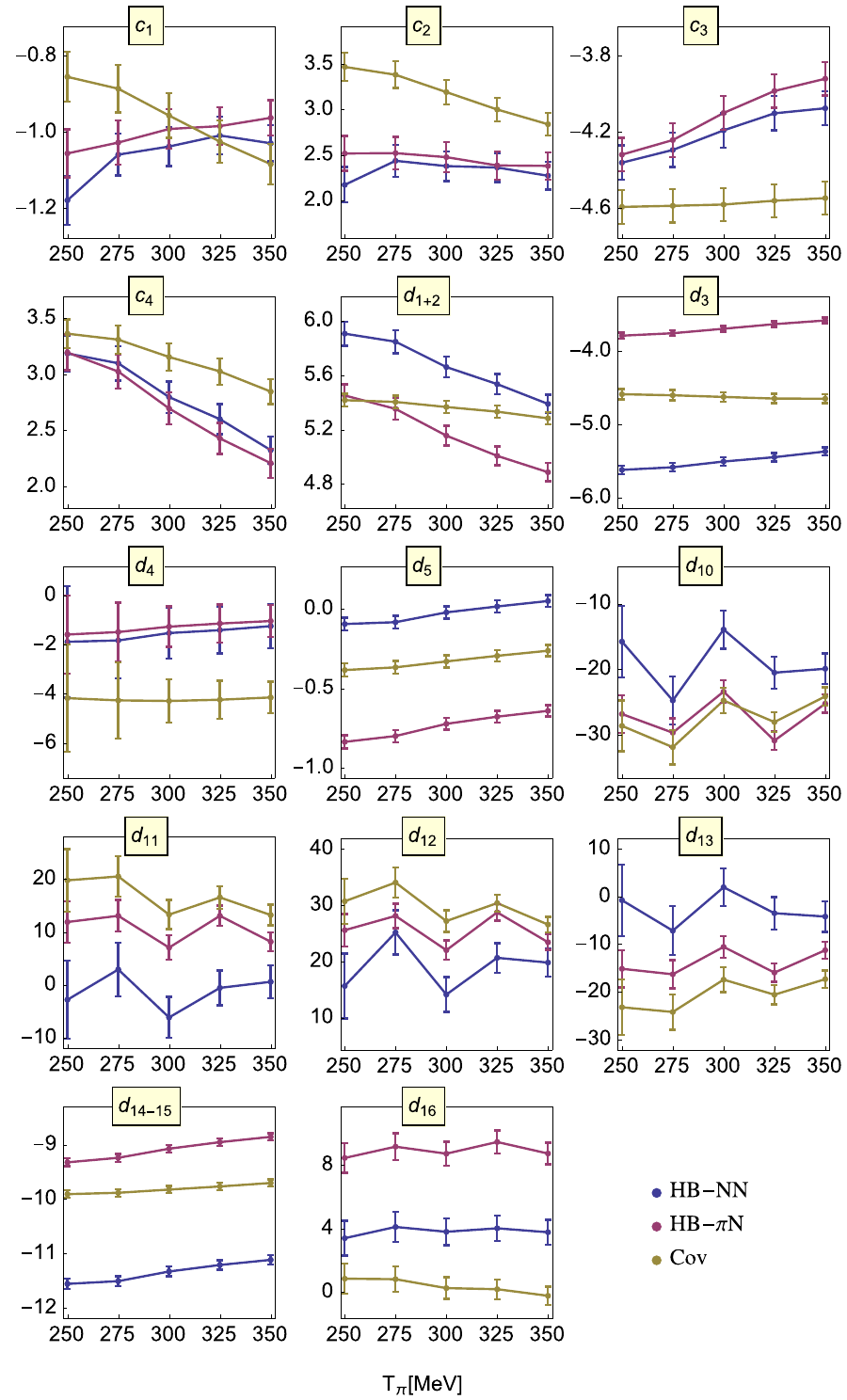}
  \caption{Change of the LECs at $Q^4$ over the maximum fit energy $T_{\pi,\pi\pi N}$.}
  \label{fig:LECsQ4P1}
\end{figure}

\begin{figure}[ht]
  \centering
\includegraphics[width=0.7\textwidth]{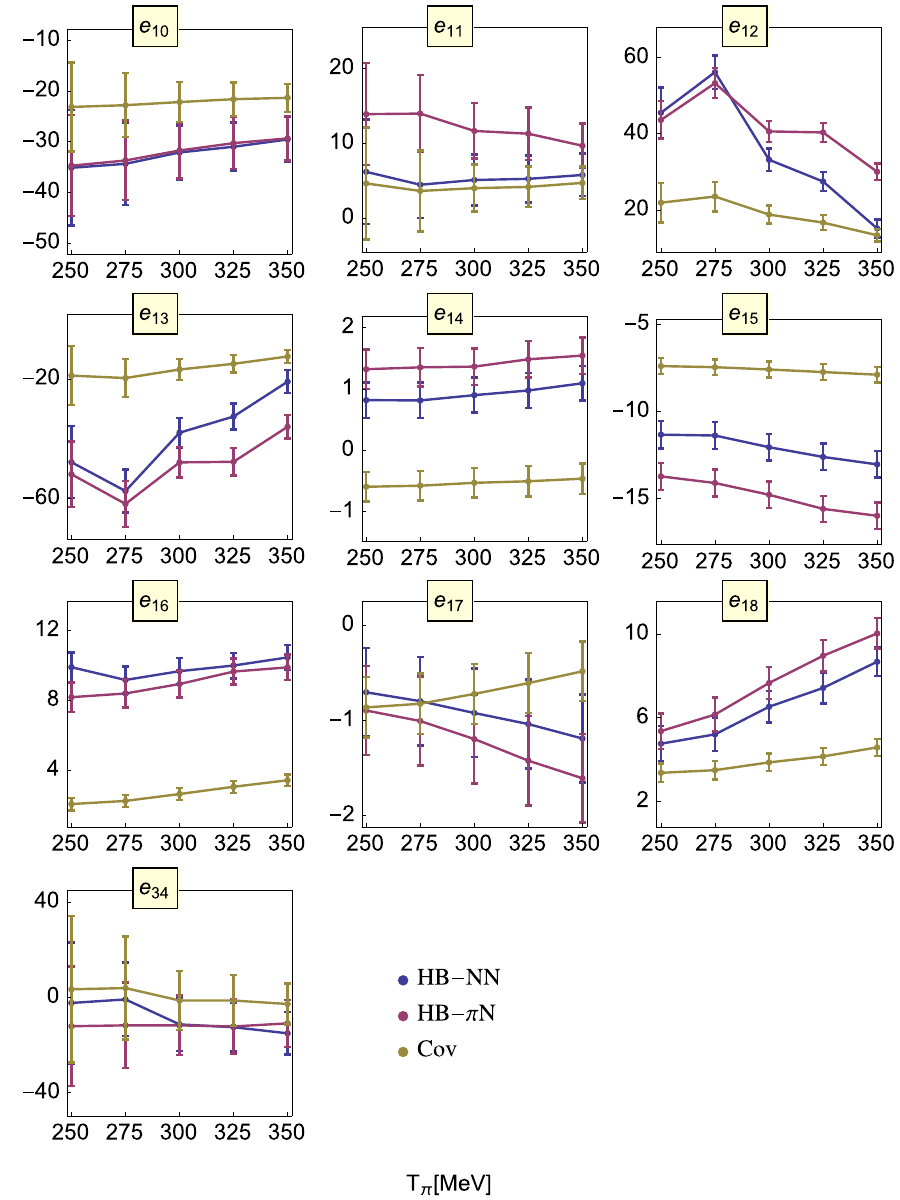}
  \caption{Change of the LECs at $Q^4$ over the maximum fit energy $T_{\pi,\pi\pi N}$.}
  \label{fig:LECsQ4P2}
\end{figure}

\newpage
\begin{figure}[ht]
\vspace{-1cm}
  \centering
  \includegraphics[width=0.65\textwidth]{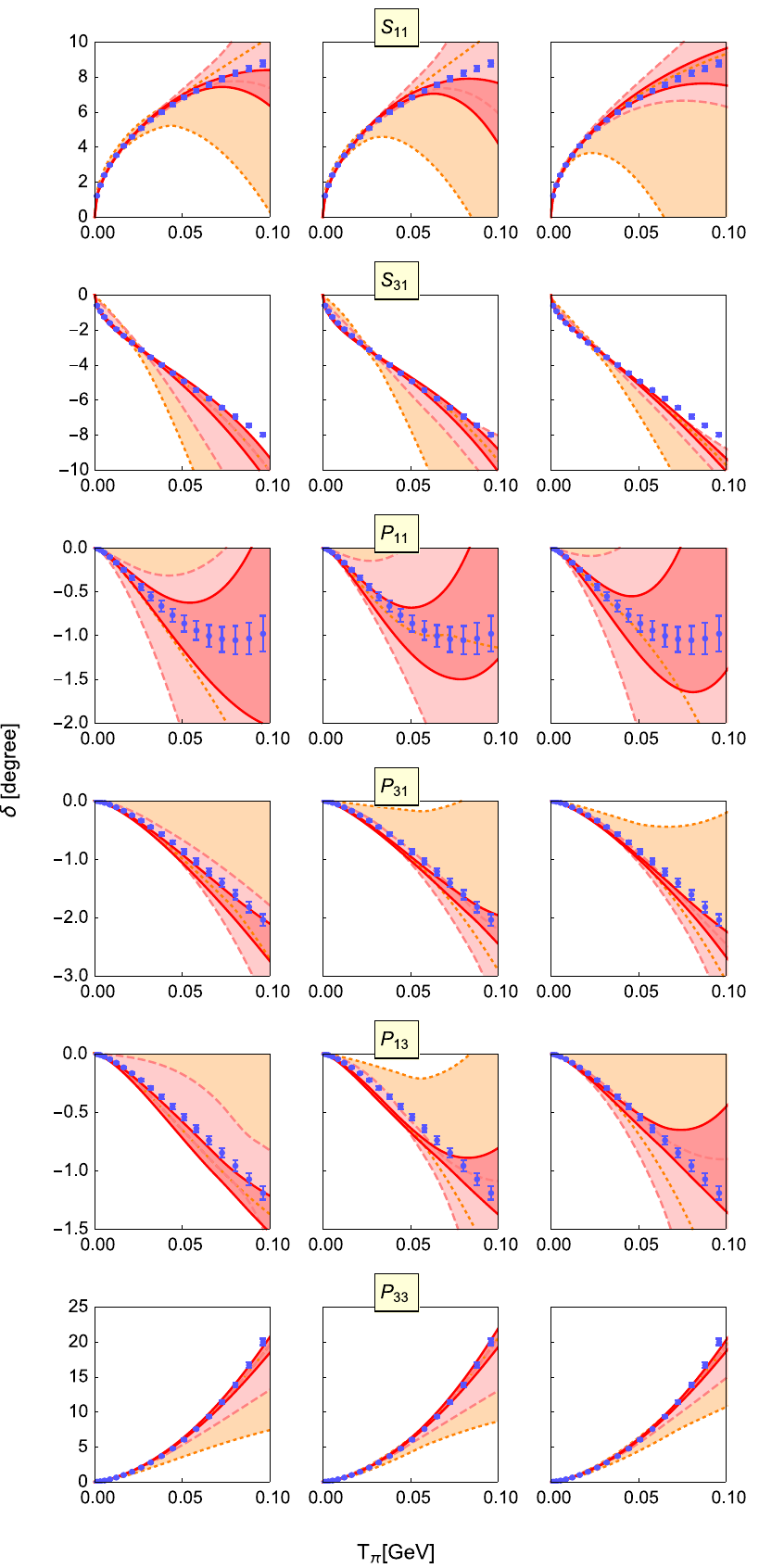}
\caption{Predictions for the $\pi N\to\pi N$ $S$ and $P$ waves up to
  $T_\pi=100$~MeV with the LECs in Table~\ref{tab:FitST} taken as input. 
  Columns from left to right correspond 
  to the predictions in the HB-NN, HB-$\pi$N and covariant counting,
  respectively.
  The orange, pink and red (dotted, dashed and solid) bands refer to 
   the $Q^2$, $Q^3$ and $Q^4$ results including theoretical uncertainties,
  respectively. }
\label{fig:SwavesST}
\end{figure}

\newpage
\begin{figure}[ht]
\vspace{-1cm}
  \centering
  \includegraphics[width=0.65\textwidth]{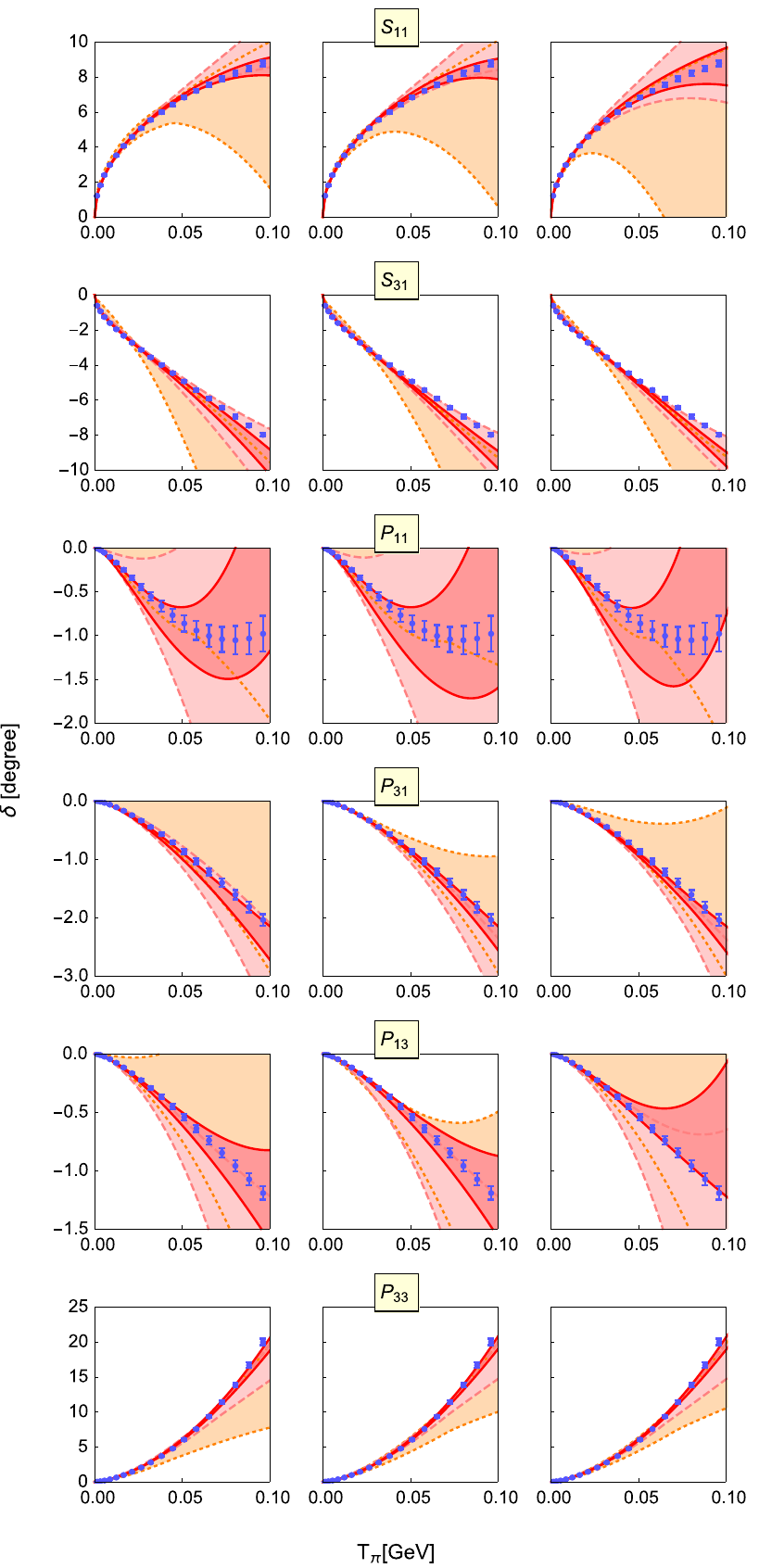}
\caption{Predictions for the $\pi N\to\pi N$ $S$ and $P$ waves up to
  $T_\pi=100$~MeV with the LECs in Tables~\ref{tab:FitpipiN1} and
  \ref{tab:FitpipiN2} taken as input. Columns from left to right correspond 
  to the predictions in the HB-NN, HB-$\pi$N and covariant counting,
  respectively. The orange, pink and red (dotted, dashed 
   and solid) bands refer to the $Q^2$, $Q^3$ and $Q^4$ results
  including theoretical uncertainties,
  respectively. }
\label{fig:Swaves}
\end{figure}

\newpage
\begin{figure}[ht]
\vspace{-1cm}
  \centering
  \includegraphics[width=0.7\textwidth]{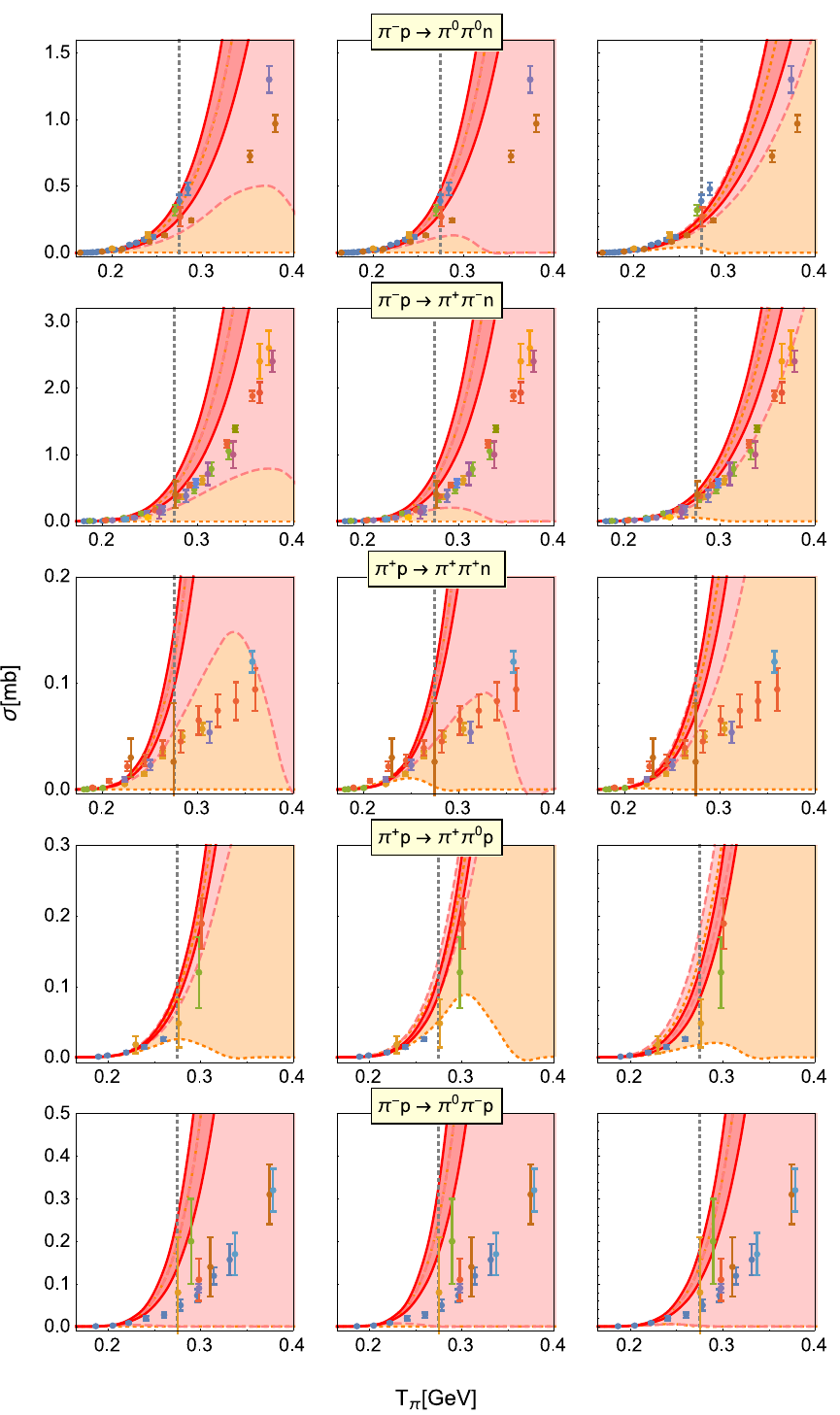}
\caption{Predictions for the $\pi N\to\pi\pi N$ total cross sections up to
  $T_\pi=\unit{400}{\mega\electronvolt}$. The energies used in the fit
  are on the left of  the vertical dotted lines. For remaining notation see the
  caption of Fig.~\ref{fig:Swaves}.}
\label{fig:sigmatot}
\end{figure}


\newpage
\begin{figure}[ht]
  \centering
\subfigure{
 \includegraphics[width=0.7\textwidth]{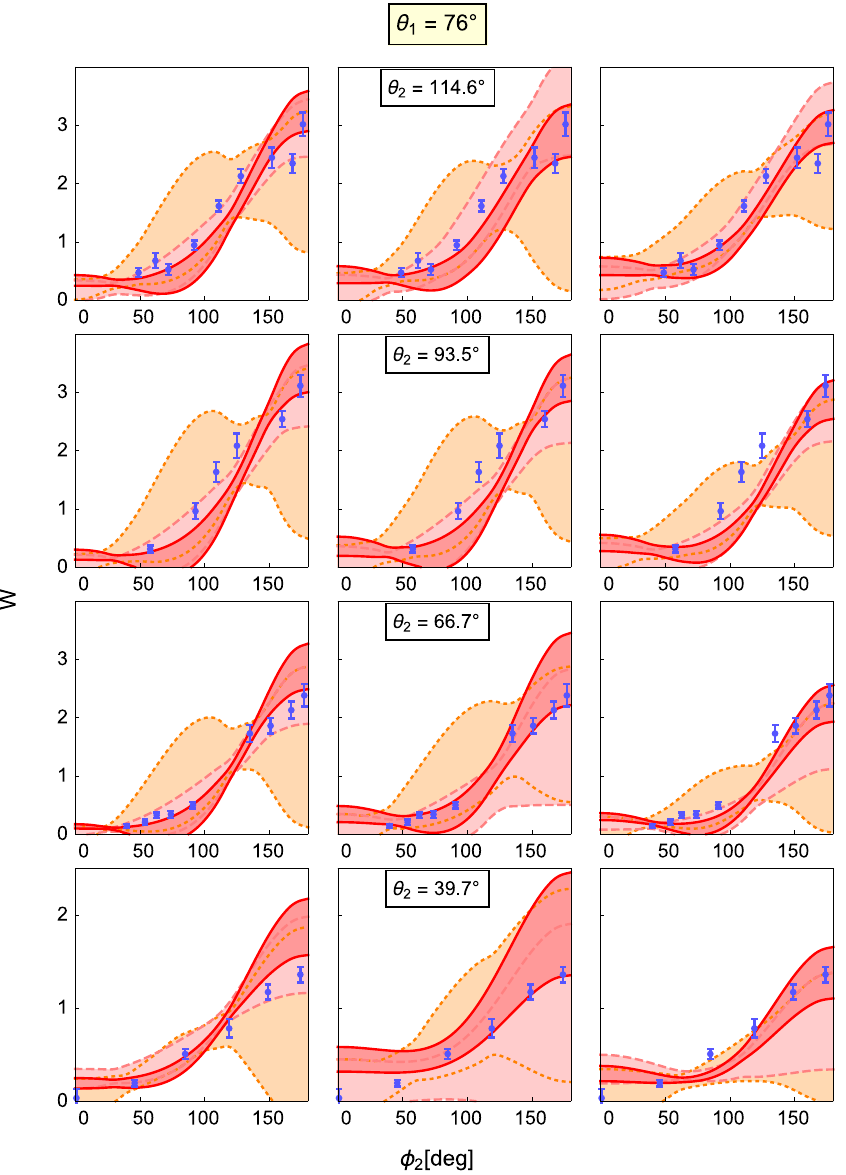}}
  \caption{Predictions for the angular correlation functions
  in the $\pi^-p\to\pi^+\pi^-n$ channel at fixed $\theta_1$ and
  $\theta_2$ for
$T_\pi=\unit{280}{\mega\electronvolt}$. 
The lower/middle/upper panel
  correspond to the HB-NN, HB-$\pi$N and covariant counting. The orange, pink and red (dotted, dashed and solid) bands refer to the $Q^2$, $Q^3$ and $Q^4$ results
  including theoretical uncertainties,
  respectively.}
  \label{fig:sigmadiffW1}
\end{figure}

\newpage
\begin{figure}[ht]
  \centering
\subfigure{
 \includegraphics[width=0.7\textwidth]{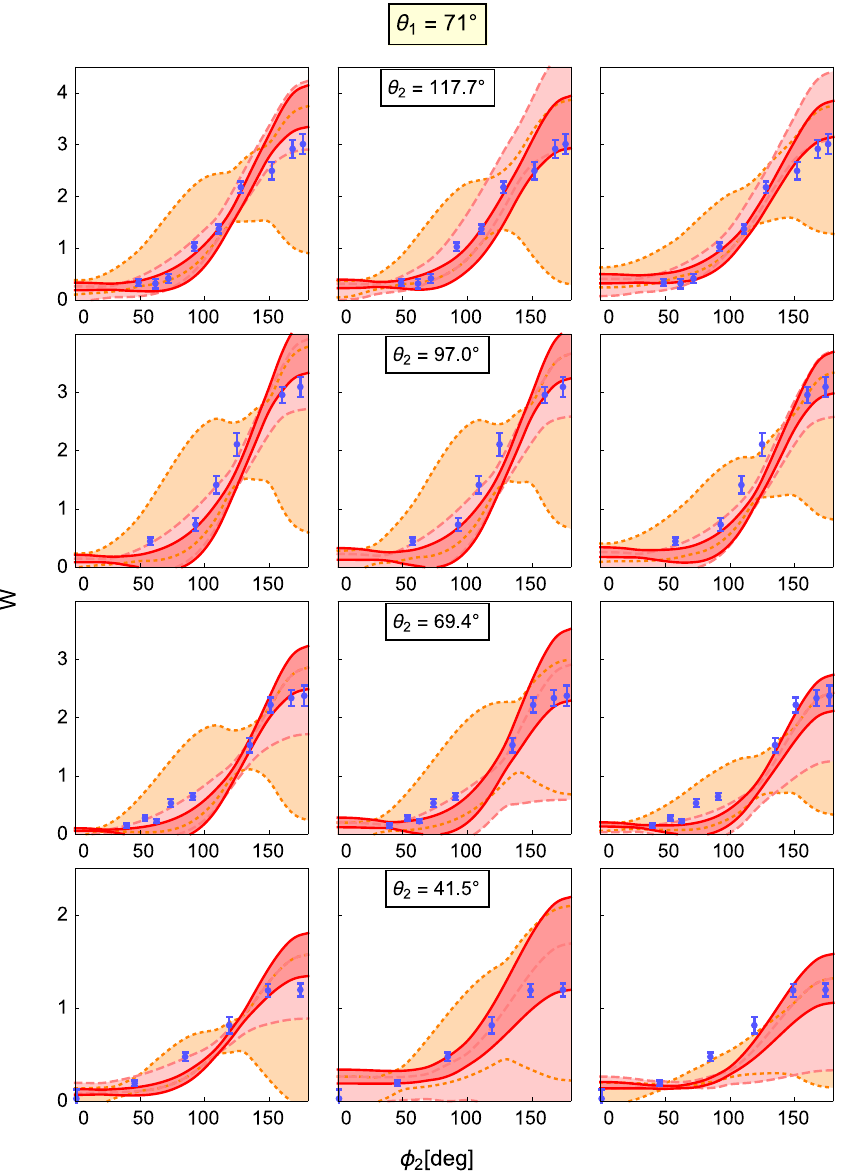}}
  \caption{Predictions for the angular correlation functions
  in the $\pi^-p\to\pi^+\pi^-n$ channel at fixed $\theta_1$ and
  $\theta_2$ for
$T_\pi=\unit{280}{\mega\electronvolt}$. 
The lower/middle/upper panel
  correspond to the HB-NN, HB-$\pi$N and covariant counting. The orange, pink and red (dotted, dashed and solid) bands refer to the $Q^2$, $Q^3$ and $Q^4$ results
  including theoretical uncertainties,
  respectively.}
  \label{fig:sigmadiffW2}
\end{figure}

\newpage
\begin{figure}[ht]
  \centering
\subfigure{
 \includegraphics[width=0.7\textwidth]{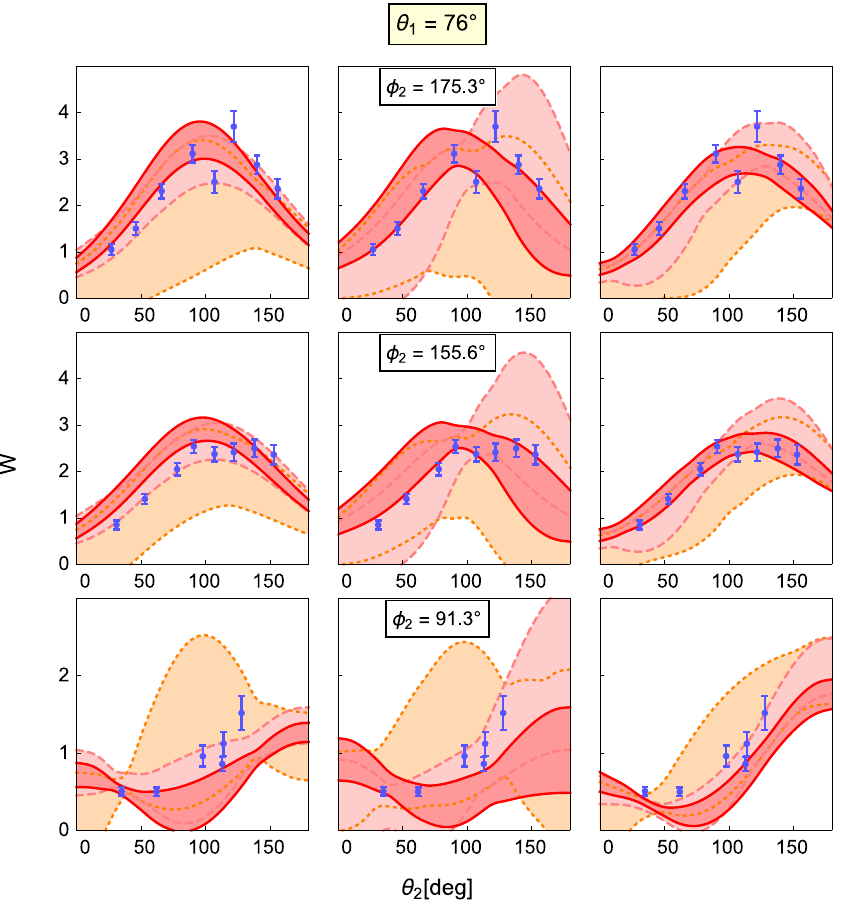}}
  \caption{Predictions for the
    angular correlation functions in the $\pi^-p\to\pi^+\pi^-n$
    channel at fixed $\theta_1$ and $\phi_2$ for
    $T_\pi=\unit{280}{\mega\electronvolt}$. For
    remaining notation see Fig.~\ref{fig:sigmadiffW1}. }
  \label{fig:sigmadiffW3}
\end{figure}
\newpage
\begin{figure}[ht]
  \centering
\subfigure{
 \includegraphics[width=0.7\textwidth]{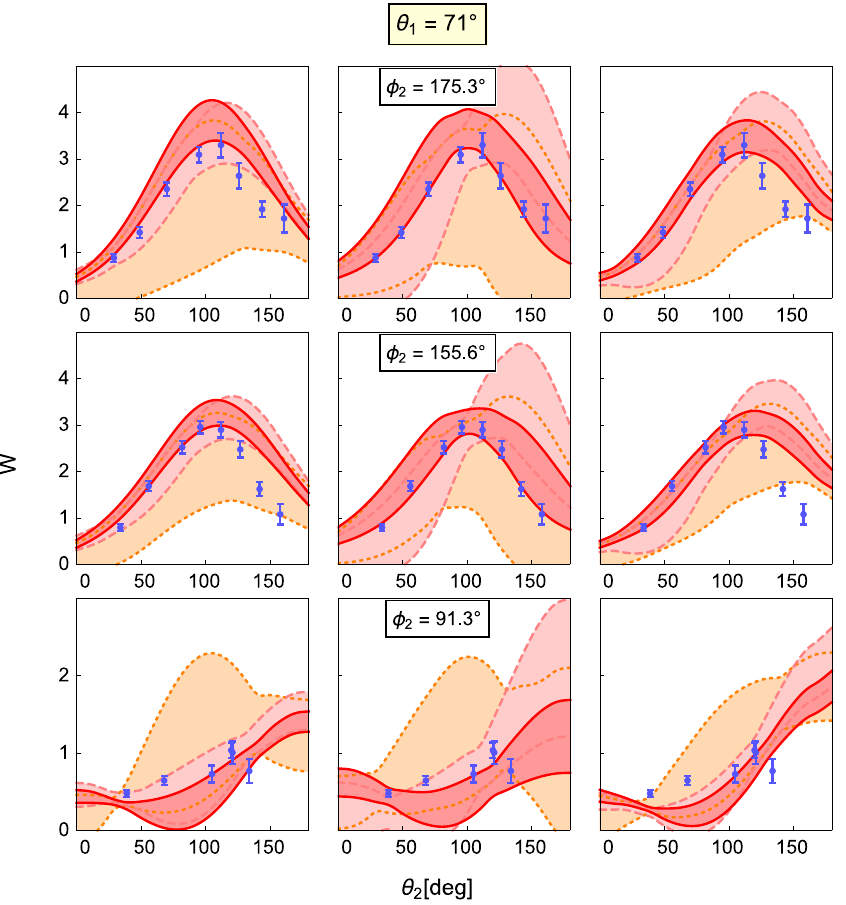}}
  \caption{Predictions for the
    angular correlation functions in the $\pi^-p\to\pi^+\pi^-n$
    channel at fixed $\theta_1$ and $\phi_2$ for
    $T_\pi=\unit{280}{\mega\electronvolt}$. For
    remaining notation see Fig.~\ref{fig:sigmadiffW1}. }
  \label{fig:sigmadiffW4}
\end{figure}

\newpage
\begin{figure}[ht]
  \centering
\subfigure{
 \includegraphics[width=0.7\textwidth]{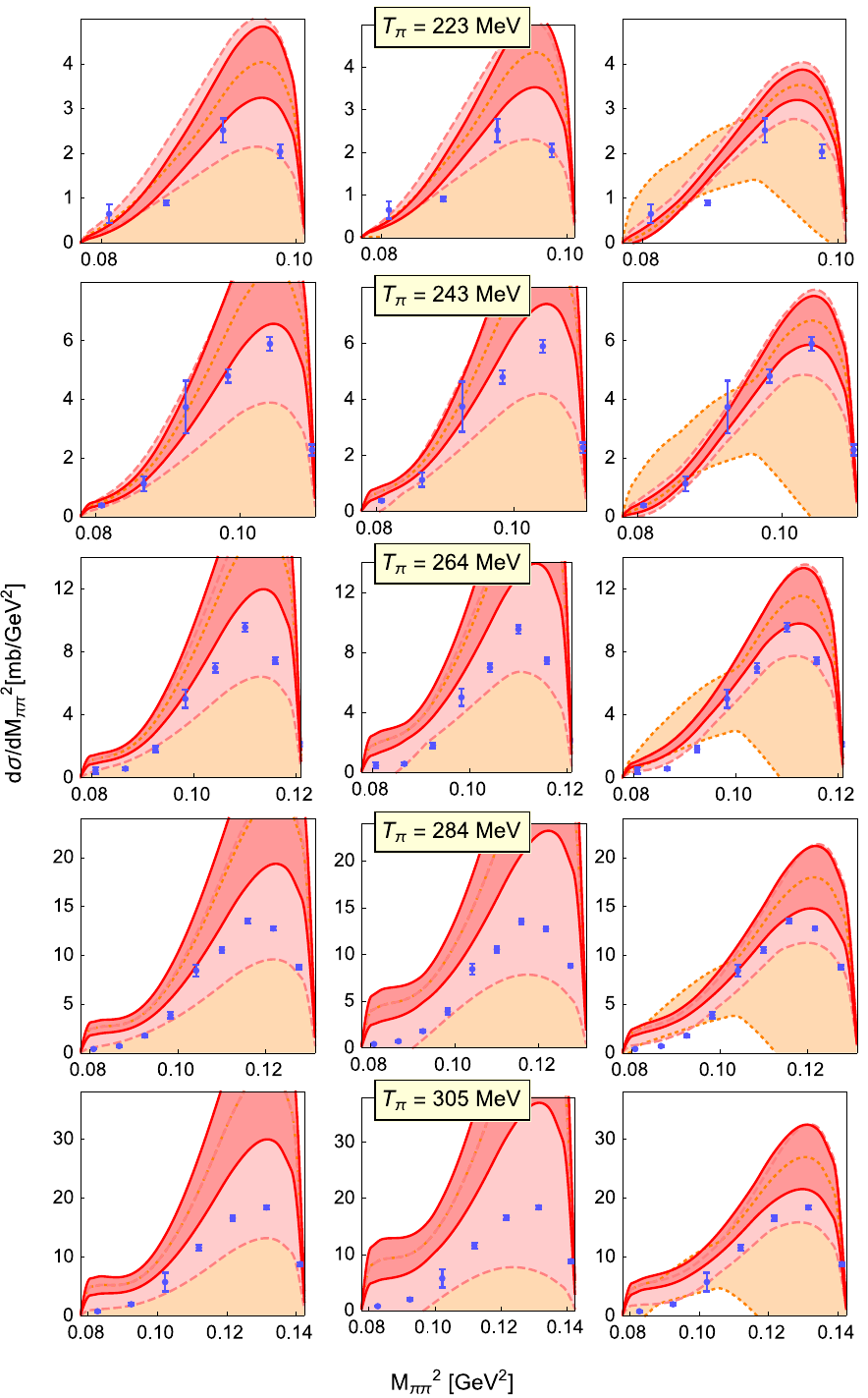}}
  \caption{Predictions 
    for the single-differential cross sections with respect to
    $M_{\pi\pi}^2$ for the channel
    $\pi^-p\to\pi^+\pi^-n$.  For
    remaining notation see Fig.~\ref{fig:Swaves}.}
  \label{fig:sigmadiffdMsq2}
\end{figure}


\newpage
\begin{figure}[ht]
  \centering
\subfigure{
 \includegraphics[width=0.7\textwidth]{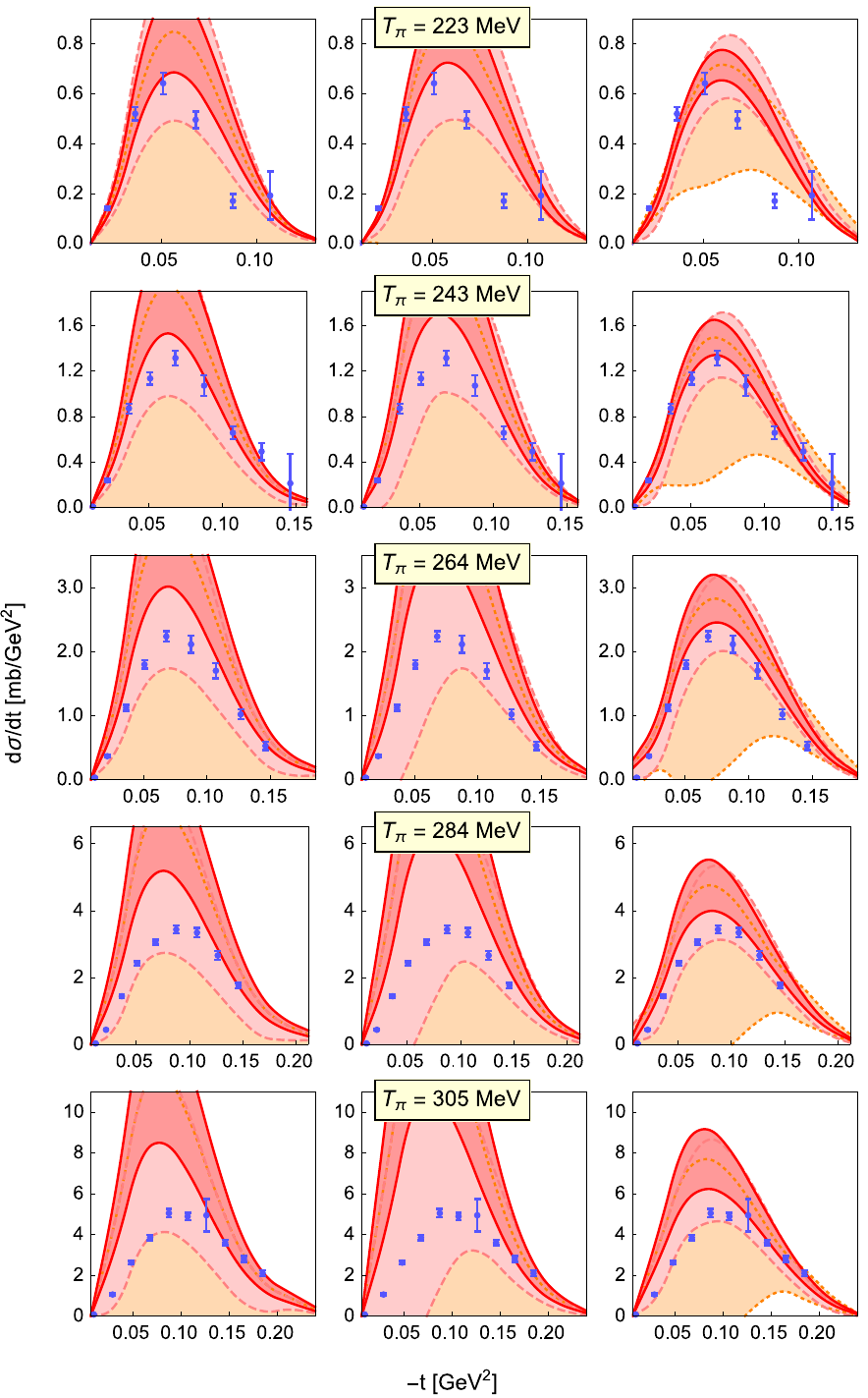}}
  \caption{Predictions 
    for the single-differential cross sections with respect to
    $t$ for the channel
    $\pi^-p\to\pi^+\pi^-n$.  For
    remaining notation see Fig.~\ref{fig:Swaves}.}
  \label{fig:sigmadiffdt2}
\end{figure}

\newpage
\begin{figure}[ht]
\vspace{-1cm}
  \centering
  \includegraphics[width=0.7\textwidth]{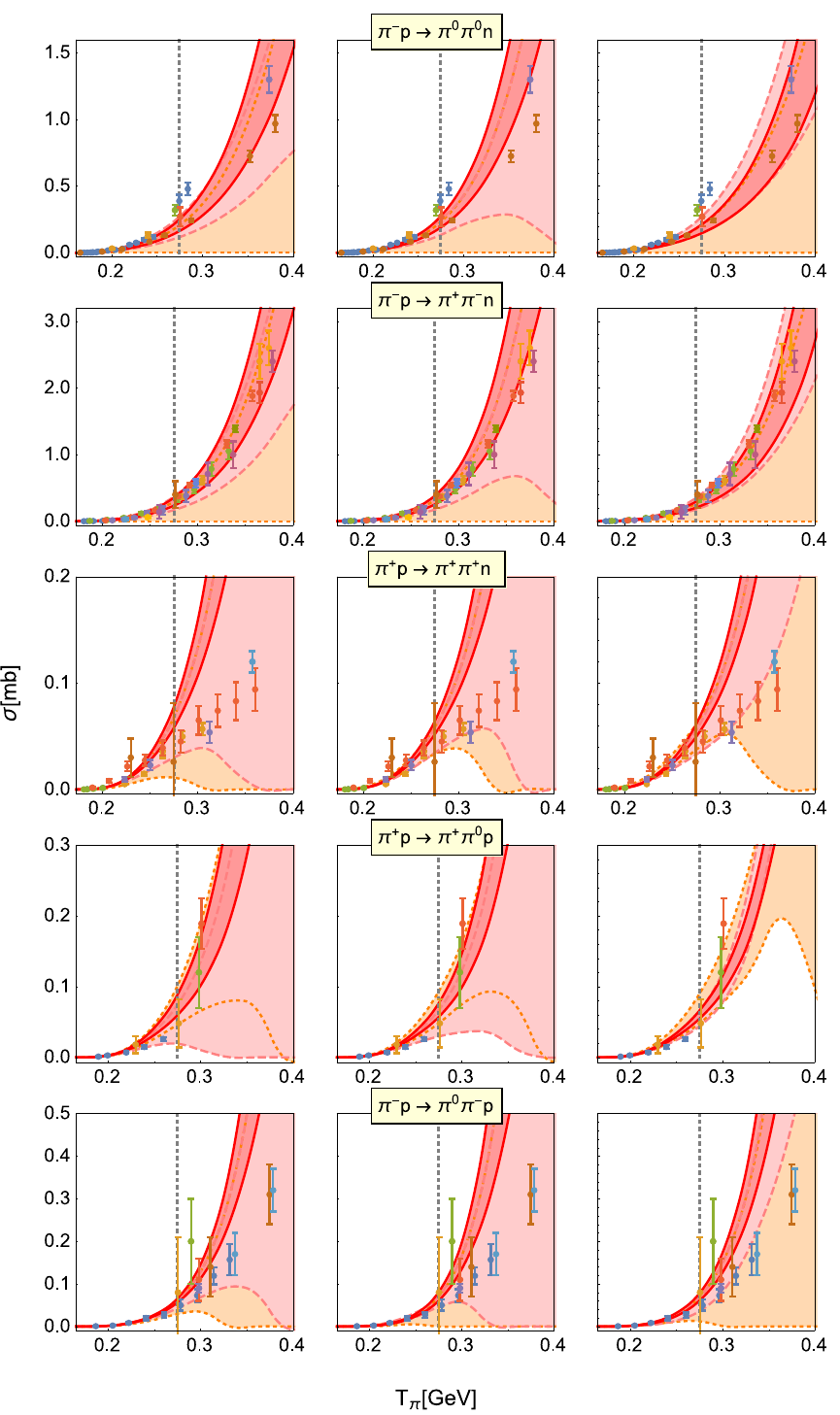}
\caption{Predictions for the $\pi N\to\pi\pi N$ total cross sections up to
  $T_\pi=\unit{400}{\mega\electronvolt}$. Columns from left to right correspond 
  to the predictions in the HB-NN, HB-$\pi$N and covariant counting,
  respectively. The orange, pink and red (dotted, dashed and solid) bands refer to the $Q^2+\delta^1$, $Q^3+\delta^1$ and $Q^4+\delta^1$ results
  including theoretical uncertainties,
  respectively. }
\label{fig:sigmatotDelta}
\end{figure}


\newpage
\begin{figure}[ht]
  \centering
\includegraphics[width=\textwidth]{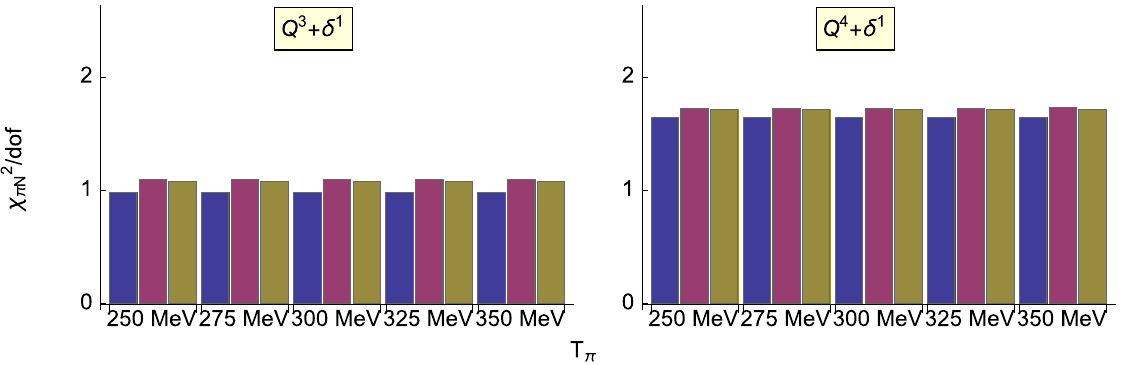}
\includegraphics[width=\textwidth]{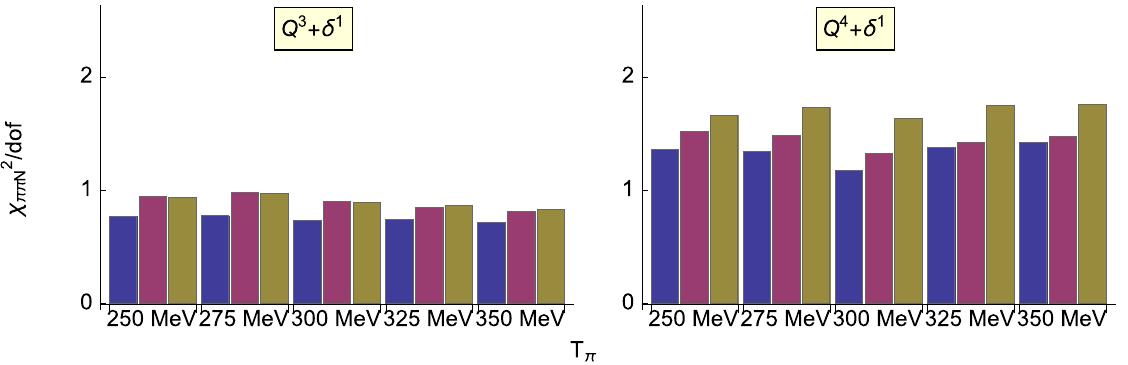}
\includegraphics[width=\textwidth]{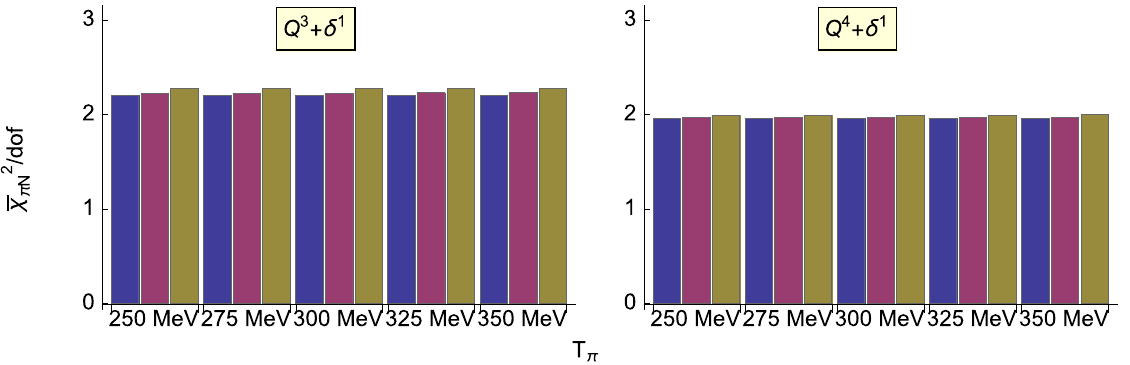}
\includegraphics[width=\textwidth]{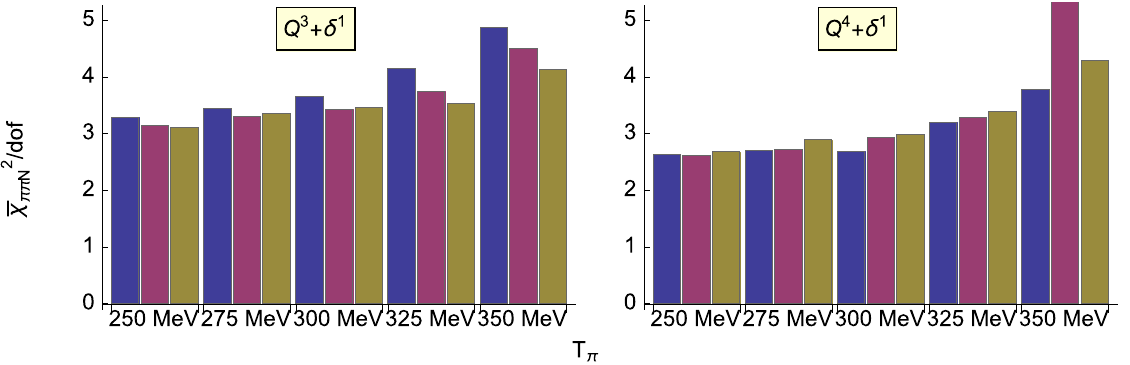}
\caption{Reduced $\chi_{\pi N}^2$/$\chi_{\pi\pi N}^2$ (with theoretical error) 
         and $\bar\chi_{\pi N}^2$/$\bar\chi_{\pi\pi N}^2$
         (without theoretical error) for fits including leading $\Delta$-pole
         contributions up to various maximum energy
         $T_{\pi,\pi\pi N}$. The blue/red/green bars denote the results for the 
          HB-NN/HB-$\pi$N/Cov counting.}
  \label{fig:RedChiSqWithDelta}
\end{figure}

\end{document}